\documentclass[a4paper,12pt]{article}
\usepackage{natbib}
\usepackage[english]{babel}

\usepackage{booktabs,dcolumn}
\usepackage{psfrag,graphicx}
\usepackage{eso-pic,graphicx}
\usepackage{amsmath}
\usepackage{graphicx}
\usepackage{pdflscape}
\usepackage{longtable}
\usepackage{epstopdf}
\usepackage{appendix}
\usepackage{dashbox}
\usepackage{caption}

\newcolumntype{z}[1]{D{.}{.}{#1}}

\date{}

\begin{document}

\title{
\begin{center}
{\Large \bf Nonparametric Expected Shortfall Forecasting Incorporating Weighted Quantiles} \end{center}
}
\author{Giuseppe Storti$^{1}$, Chao Wang$^{2}$
\\
$^{1}$ Department of Economics and Statistics, University of Salerno\\
$^{2}$ Discipline of Business Analytics, The University of Sydney}
\date{} \maketitle

\begin{abstract}
\noindent

A new semi-parametric Expected Shortfall (ES) estimation and forecasting framework is proposed. The proposed approach is based on a two-step estimation procedure. The first step involves the estimation of Value-at-Risk (VaR) at different quantile levels through a set of quantile time series regressions. Then, the ES is computed as a weighted average of the estimated quantiles. The quantiles weighting structure is parsimoniously parameterized by means of a Beta weight function whose coefficients are optimized by minimizing a joint VaR and ES loss function of the Fissler-Ziegel class. The properties of the proposed approach are first evaluated with an extensive simulation study using two data generating processes. Two forecasting studies with different out-of-sample sizes are then conducted, one of which focuses on the 2008 Global Financial Crisis (GFC) period. The proposed models are applied to 7 stock market indices and their forecasting performances are compared to those of a range of parametric, non-parametric and semi-parametric models, including GARCH, Conditional AutoRegressive Expectile (CARE), joint VaR and ES quantile regression models and simple average of quantiles. The results of the forecasting experiments provide clear evidence in support of proposed models.

%Value-at-Risk and Expected Shortfall forecasting results favor the proposed models.

\vspace{0.5cm}

\end{abstract}

\newpage
\pagenumbering{arabic}

{\centering
\section{\normalsize INTRODUCTION}\label{introduction_sec}
\par
}
\noindent

Value-at-Risk (VaR) is employed by many financial institutions as an important risk management tool. Representing the market risk as one number, VaR has been used as a standard risk measurement metric for the past two decades. However, as recently recognized by the Basel Committee for Banking Supervision,
VaR suffers from a number of weaknesses affecting its reliability as a reference metric for determining regulatory capital requirements \citep{BIS2013}. First, VaR cannot measure the expected loss for extreme (violating) returns. In addition, it can be shown that VaR is not always a \emph{coherent} risk measure, due to failure to match the \emph{subadditivity} property. For these reasons, the Committee proposed in May 2012 to replace VaR with the Expected Shortfall (ES, \citealt{artzner1997,artzner1999}). ES is defined as the expectation of the return conditional on it having exceeded the VaR. Differently from VaR, it is a  coherent measure and it ``measures the riskiness of a position by considering both the size and the likelihood of losses above a certain confidence level'' \citep{BIS2013}. Thus, in recent years ES has been increasingly employed for tail risk measurement. However, still there is much less existing research on modeling ES compared with VaR.

The Basel III Accord, which was implemented in 2019, places new emphasis on ES. Its recommendations for market risk management are illustrated in the 2019 document ``Minimum capital requirements for market risk'' that, on page 89, mentions: ``ES must be computed on a daily basis for the bank-wide internal models to determine market risk capital requirements. ES must also be computed on a daily basis for each trading desk that uses the internal models approach (IMA).''  \citep{BIS2019}. According to the same source, in calculating ES, banks must refer to the average loss in the tail below the 2.5\% quantile level. Therefore, in the empirical application of our paper, we focus on one-step-ahead tail risk forecasting at the same level.

The literature on ES modelling and forecasting is closely related to previous research on VaR. The quantile regression type model, e.g. the Conditional Autoregressive Value-at-Risk (CAViaR) model of \cite{caviar}, is a popular semi-parametric approach to estimate and forecast VaR.

However, CAViaR type models cannot directly estimate and forecast ES. \cite{tayl2017} proposes a joint ES and quantile regression framework (ES-CAViaR) which employs the Asymmetric Laplace (AL) density to build a likelihood function whose Maximum Likelihood Estimates (MLEs) coincide with those obtained by minimisation of a strictly consistent joint loss function for VaR and ES. The frameworks in \cite{tayl2017} assume that the difference or ratio between VaR and ES follow specific dynamics, also in order to guarantee that VaR and ES do not cross with each other. Essentially, this implies additional assumptions on ES dynamics.

\cite{Fissler2016} develop a family of joint loss functions (or ``scoring functions'') that are strictly consistent for the true VaR and ES, i.e. they are uniquely minimized by the true VaR and ES series. Under specific choices of the functions involved in the joint loss function of \cite{Fissler2016}, it can be shown that the negative of AL log-likelihood function, presented in \cite{tayl2017}, can be derived as a special case of the \cite{Fissler2016} class of loss functions. \cite{pattonetal2019} propose new dynamic models for VaR and ES, through adopting the generalized autoregressive score (GAS) framework (\citealt{crealetal2013} and \citealt{harvey2013}) and utilizing the loss functions in \cite{Fissler2016}.

%There is a paper by Leorato, Peracchi and Tanase, CSDA - it's in the literature folder - that proposes
% an approach somewhat related to ours (if you remember we discussed this in Sydney). Should we cite it highlighting the main differences and advantages (if any) of our approach? Or make things simpler and omit it in the review?....What do you think?
% I BETTER THOUGHT ABOUT THIS POINT. IT IS PROBABLY TOO COMPLICATED TO INSERT THIS REFERENCE INTO THE DISCUSSION. I'LL EXPLAIN: IF WE DO IT, WE SHOULD SPEND HALF OF THE INTRODUCTION TRY TO EXPLAIN THE ADDITIONAL CONTRIBUTION WITH RESPECT TO THE PAPER BY PERACCHI AND TANASE. THIS COULD BE MISLEADING FOR A SUPERFICIAL REFEREE. ALSO, WE ARE NOT OVERLAPPING WITH THEIR WORK. IN COMMON WITH THEM WE ONLY HAVE THE IDEA OF USING THE DEFINITION OF ES AS A WEIGHTED AVERAGE OF QUANTILES. IF A REFEREE SHOULD ASK WE COULD EASILY DEMONSTRATE THE NOVELTY OF OUR APPROACH. SO I WOULD OMIT THE REFERENCE FOR NOW. AGAIN, WHAT DO YOU THINK?

In our paper, a new ES estimation and forecasting framework is proposed where the ES is modelled as an affine function of tail quantiles. Hence, we refer to our approach as the \emph{Weighted Quantile} estimator. The quantiles are produced from the CAViaR model of \cite{caviar} by grid search of a range of equally spaced quantile levels below the target VaR level, i.e. 2.5\%. We will discuss the selection details of these quantile levels later. For large grid sizes, the weighting pattern of the selected quantiles is based on a two parameter Beta weight function. The Beta weight function is a parsimonious but yet flexible choice and is able to reproduce a variety of different behaviours such as declining, increasing or hump shaped patterns. For less dense grids, direct estimation of the weights can be also entertained.
%lso called the "Beta lag", borrowed from the literature on Mixed Data Sampling (MIDAS, \citealt{ghysels2007midas}) regression models.

We estimate the parameters of the Beta weight function, determining the individual weights assigned to the selected quantiles, by minimizing strictly consistent VaR and ES joint loss functions of the class defined in \cite{Fissler2016}. In particular we focus on the AL loss in \cite{tayl2017}.

It is worth noting that the proposed estimator does not require any additional assumption on the dynamics of the ES process, but it only relies on the natural definition of ES as the tail expectation of the conditional distribution of returns\footnote{Under the assumption that this distribution is continuous, as later explained.}, so reducing model uncertainty and risk of potential mis-specification on the ES.

%\textbf{ Therefore, our estimation procedure is able to automatically guarantee that ES is lower than VaR without the crossing issue which needs to be separately addressed  in \cite{tayl2017}}.
% that can be explicitly derived by standard algebraic manipulations, are those naturally implied by its definition in terms of the expectation of tail-quantiles. Furthermore, it directly relies on the mathematical definition of ES.

%It follows the ES can be predicted without having to specify an additional dynamic equation

%The dynamics of the ES are those naturally implied by its definition in terms of the expectation of quantiles in the tail, which is calculated as the weighted average of quantiles that are below the 2.5\% quantile level.

%Second, it is important to emphasize that we derive the ES without specifying its additional dynamics, aiming to reduce the model uncertainty or potential mis-specification on the ES side.

Our method has some interesting connections with the existing literature. First, there are some evident affinities between our method and the Conditional Autoregressive Expectile (CARE) models proposed by \cite{tayl2008}. These are semi-parametric models that directly estimate quantiles and expectiles, and implicitly ES, through a set of expectile regressions \citep{newey1987}. To select the appropriate expectile levels for VaR and ES estimation, implementation of CARE type models requires a grid search process which is relatively computationally expensive (dependent on the complexity of the model and the size of the grid).
Namely, both our framework and CARE models involve a two-step estimation procedure and a grid search process. Later, we will show that our framework can produce more accurate ES forecasting results than CARE. %First, it is closely related to literature on forecasts combination. \cite{tayl2019} has recently proposed to use a forecast combination of different VaR\&ES models of the same order. However, our strategy is substantially different since we are combining forecasts of a list of VaR models (CAViaR) of different quantile orders, instead of a list of different models.
Further, the proposed framework has also some connections with literature on forecasts combination. \cite{taylor2020forecast} has recently proposed to use a forecast combination of different VaR and ES models of the same order. However, our strategy differs from the one adopted by \cite{taylor2020forecast} since we are combining forecasts from a list of VaR models (CAViaR) of different quantile orders, instead of a list of different joint VaR and ES models.

%Third, another nice feature of our approach is that it simplifies the introduction of external regressors in the ES dynamics. Once the regressors are included in the VaR model, the weighted average naturally extends their effect to the ES model, with no need for the estimation of additional parameters (to be checked and discussed together).\textbf{=> Maybe put into the conclusion and future work section I AGREE,  LET'S EVENTUALLY MOVE THIS TO THE CONCLUDING REMARKS SECTION}.

The paper is organized as follows. Section \ref{model_review_section} presents a  review of the ES-CAViaR and CARE models while Section \ref{model_section} formalizes and discusses the proposed weighted quantile framework. Further, we present empirical evidence on both simulated and real market index data. First, Section \ref{simulation_section} presents the results of a Monte Carlo simulation aimed at providing an appraisal of the finite-sample statistical properties of the proposed estimators and of their sensitivity to the implementation settings, that is to the choice of the size and the lower bound of the grid. Section \ref{data_empirical_section}, then, assesses the effectiveness of the proposed framework in standard risk management applications by presenting the results of two out-of-sample tail risk forecasting exercises, in which the performances of the weighted quantile estimators are compared with those of some state-of-the-art competitors. Section \ref{conclusion_section} concludes the paper and discusses future work.\\

{\centering
\section{\normalsize JOINT MODELLING OF VaR AND ES: ES-CAViaR AND CARE MODELS}\label{model_review_section}
\par
}

%\subsection{Models based on strictly consistent scoring functions}

%\subsection{General framework}

To start, let $\mathcal{I}_t$ be the information available at time $t$ and
\[
F_{t}(r)=Pr(r_{t}\leq r | \mathcal{I}_{t-1})
\]
be the Cumulative Distribution Function (CDF) of return $r_{t}$ conditional on $\mathcal{I}_{t-1}$. We assume that $F_{t}(.)$ is strictly increasing and continuous on the real line $\Re$. Under this assumption, the one-step-ahead $\alpha$ level Value at Risk at time $t$ can be defined as
\[
Q_{t,\alpha}=F^{-1}_{t}(\alpha)\qquad 0 <\alpha <1.
\]
Within the same framework, the one-step-ahead $\alpha$ level  Expected Shortfall can be shown \citep[see][among others]{AceTas2002} to be equal to the tail conditional expectation of $r_t$
\begin{equation}
    ES_{t,\alpha}=E(r_t|r_t\leq Q_{t,\alpha}, \mathcal{I}_{t-1}).
\label{e:ESdef}
\end{equation}
In the literature, several models for the estimation and prediction of VaR and ES have been proposed. In this section we present a review of the main approaches focusing on estimation approaches based on the minimization of strictly consistent scoring functions.  In order to simplify notation, in the remainder, unless differently specified, the following notational conventions are adopted: $\text{ES}_{t,\alpha}\equiv ES_{t}$ and $Q_{t,\alpha} \equiv Q_t$, where $\alpha$ denotes the target level for the estimation of VaR and ES (in the empirical application we focus on $\alpha=0.025$).

%\subsection{Models based on strictly consistent scoring functions}

\cite{koenkermachado1999} show that the quantile regression estimator is equivalent to a maximum likelihood estimator when assuming that the data are conditionally distributed as an Asymmetric Laplace (AL) with a mode at the quantile of interest. If $r_t$ is the return on day $t$ and $Pr(r_t < Q_t | \mathcal{I}_{t-1}) = \alpha$, then the parameters in the model for $Q_t$ can be estimated maximizing a quasi-likelihood based on:
$$ p(r_t| \mathcal{I}_{t-1}) = \frac{\alpha (1-\alpha)}{\sigma} \exp \left( -(r_t-Q_t)(\alpha - I(r_t < Q_t) )/ \sigma  \right), $$
for $t=1,\ldots,N$ and where $\sigma$ is a scale parameter.

\cite{tayl2017} extends this result to incorporate the associated ES quantity into the likelihood expression, noting a link between $\text{ES}_t$ and a dynamic $\sigma_t$, resulting in the conditional density function:
\begin{eqnarray} \label{es_var_likelihood}
p(r_t| \mathcal{I}_{t-1}) = \frac{(\alpha - 1)}{ES_t} \exp \left( \frac{(r_t-Q_t)(\alpha - I(r_t < Q_t)) }{\alpha ES_t}  \right).
\end{eqnarray}
This allows a likelihood function to be built and maximised, given model expressions for $(Q_t, \, ES_t)$. In Eq. (\ref{es_var_likelihood}),  $r_{t}$ is the daily return, $Q_{t}$ and $\text{ES}_{t}$ denote the $\alpha$ target level VaR and ES on day $t$. \cite{tayl2017} notes that the negative logarithm of the resulting likelihood function is strictly consistent
for $(Q_t,\, ES_t)$ considered jointly, e.g. it fits into the class of jointly consistent scoring functions for VaR and ES developed by \cite{Fissler2016}.

Members of this family are strictly consistent for $(Q_t,ES_t)$, i.e. their expectations are uniquely minimized by the true VaR and ES series. The general form of this functional family is:
\begin{eqnarray}
S_t(r_t, Q_t, ES_t) &=& (I_t -\alpha)G_1(Q_t) - I_tG_1(r_t) +  G_2(ES_t)\left(ES_t-Q_t + \frac{I_t}{\alpha}(Q_t-r_t)\right) \nonumber\\
                      &-& H(ES_t) + a(r_t) \, ,
                      \label{e:fzloss}
\end{eqnarray}
where $I_t=1$ if $r_t<Q_t$ and 0 otherwise, for $t=1,\ldots,N$, $G_1(.)$ is increasing, $G_2(.)$ is strictly increasing and strictly convex,
$G_2 = H^{'}$ and $\lim_{x\to -\infty} G_2(x) = 0$ and $a(\cdot)$ is a real-valued integrable function.

As discussed in \cite{tayl2017}, assuming $r_t$ to have zero mean, making the choices: $G_1(x) =0$,
$G_2(x) = -1/x$, $H(x)= -\text{log}(-x)$ and  $a= 1-\text{log} (1-\alpha)$, which satisfy the required criteria, returns the scoring function:
\begin{eqnarray}\label{es_caviar_log_score}
S_t(r_t, Q_t, ES_t) = -\text{log} \left( \frac{\alpha-1}{\text{ES}_t} \right) - {\frac{(r_t-Q_t)(\alpha-I(r_t\leq Q_t))}{\alpha \text{ES}_t}},
\end{eqnarray}
where the aggregated loss is indicates as $\mathbf{S} = \sum_{t=1}^N S_t$. \cite{tayl2017} refers to Eq. (\ref{es_caviar_log_score}) as the AL log score which is a strictly consistent scoring function that is jointly minimized by the true VaR and ES series. The negative of Eq. (\ref{es_caviar_log_score}) equals to the log of Eq. (\ref{es_var_likelihood}) and can be treated as the AL log-likelihood.

Within this framework, \cite{tayl2017} proposes joint models for the dynamics of VaR and ES. The proposed approach avoids ES estimates crossing the corresponding VaR estimates by combining a CAViaR model for the VaR with a choice two different specifications for the ES component. The proposed models are presented below in Eq. (\ref{es_caviar_add_model}) (ES-CAViaR-Add: ES-CAViaR with an additive VaR to ES component) and Eq. (\ref{es_caviar_mult_model}) (ES-CAViaR-Mult: ES-CAViaR with a multiplicative VaR to ES component).

\noindent \textbf{ES-CAViaR-Add:}

\begin{eqnarray} \label{es_caviar_add_model}
Q_{t}&=& \beta_1+ \beta_2 |r_{t-1}| + \beta_3 Q_{t-1},\\ \nonumber
 \text{ES}_t&=&Q_t-x_t, \\ \nonumber
 x_t&=&
\begin{cases}
    \gamma_0 + \gamma_1 (Q_{t-1} - r_{t-1}) + \gamma_2 x_{t-1} & \text{if } r_{t-1} \leq Q_{t-1},\\
    x_{t-1}              & \text{otherwise},
\end{cases}
\end{eqnarray}
where $\beta_1, \beta_2, \beta_3, \gamma_0, \gamma_1, \text{and } \gamma_2$ are parameters to be estimated.  To ensure that the VaR and ES series do not cross, \cite{tayl2017} imposes the following constraints: $\gamma_0 \ge 0, \gamma_1 \ge 0, \gamma_2 \ge 0$.\\

\noindent \textbf{ES-CAViaR-Mult:}

\begin{eqnarray} \label{es_caviar_mult_model}
Q_{t}&=& \beta_1+ \beta_2 |r_{t-1}| + \beta_3 Q_{t-1}, \\ \nonumber
 \text{ES}_t&=& \left( 1+\exp(\gamma_0) \right) Q_t, \nonumber
\end{eqnarray}
where $\beta_1, \beta_2, \beta_3, \text{and } \gamma_0$ are parameters to be estimated. $\gamma_0$ is unconstrained.

%The joint VaR \& ES loss function study is conducted to compare the models VaR and ES forecasts jointly, and to help clarify and quantify any
%extra efficiency can be gained from the Realized(-Threshold)-ES-CARE ES forecasts compared to its competitors.

%\subsection{Models based on expectiles}

A different approach to joint estimation of VaR and ES is based on the theory of expectiles. The concept of expectile is closely related to the concept of quantile. The $\tau$ level expectile $\mu_{\tau}$, as defined by \cite{aigneretal1976}, can be estimated through minimizing the following Asymmetric Least Squares (ALS) criterion \citep{newey1987}:
\begin{equation}\label{als_equation}
\sum_{t=1}^{N} |\tau-I(r_t<\mu_{\tau})|(r_t-\mu_{\tau})^2  \, ,
\end{equation}
no distributional assumption is required to estimate $\mu_{\tau}$ here.

As discussed in Section \ref{introduction_sec}, the $\alpha$ level conditional ES is defined as $\text{ES}_{t}= E(r_t|r_t\leq Q_{t},\mathcal{I}_{t-1})$.
%which stands for the expected
%value of $Y$, conditional on the set of $Y$ that is more extreme than the $\alpha$-level quantile of Y, denoted $Q_{\alpha}$.
\cite{tayl2008} shows that this is related to the conditional $\tau$ level expectile $\mu_{t,\tau}$ by the  relationship:
\begin{equation}\label{expectile_es_equation}
ES_{t}= \left(1+\frac{\tau}{(1-2\tau)\alpha} \right)\mu_{t,\tau} \, ,
\end{equation}
where $\mu_{t,\tau}=Q_{t}$, i.e. $\tau$ level expectile $\mu_{t,\tau}$ occurs at the $\alpha$ level quantile of $r_t$. Thus, $\mu_{t;\tau}$ can be used
to estimate the $\alpha$ level conditional quantile $Q_{t}$ and then scaled to estimate the associated $\text{ES}_{t}$.

Exploiting this relationship, \cite{tayl2008} proposes the CARE type models which have a similar form to the CAViaR models of \cite{caviar}, where lagged returns drive the expectiles and model parameters are estimated via minimizing an ALS criterion. The general Symmetric Absolute Value (SAV) form of this model is:

\noindent
\textbf{CARE-SAV}:
\begin{align*}
 \mu_{t,\tau} = \beta_{1} + \beta_{2} |r_{t-1}| + \beta_{3} \mu_{t-1,\tau}
\end{align*}
\noindent
where $\beta_1, \beta_2, \text{and } \beta_3$ are the parameters to be estimated, and $\mu_{t,\tau}$ is the $\tau$ level expectile on day $t$. The CARE-type model produces one-step-ahead forecasts of expectiles ($\mu_{t,\tau}$), that can be employed as VaR estimates ($Q_{t}$), by an appropriate choice of $\tau$. The VaR estimates can be further scaled, using Eq. (\ref{expectile_es_equation}), to produce forecasts of ES which cannot be directly calculated under the CAViaR framework.

However, the selection of the appropriate expectile level $\tau$ requires a grid search, based on the optimization of the violation rate (VRate, the percentage of returns exceeding VaR estimates) or of the aggregated quantile loss function  \citep{gerlach2020recare}.
%\blue{\textbf{What about cutting the text in blue to make this section shorter as asked by the referee?} Specifically, in the first case, for each grid value of $\tau$, the ALS estimator of the CARE equation parameters $\beta_j$ ($j=1, 2, 3$) is found, yielding an associated VRate($\tau$). $\hat{\tau}$ is then set to the grid value of $\tau$ s.t. VRate is closest to the desired $\alpha_\tau$. Differently, when the aggregated quantile loss is chosen as an objective function, the selected $\alpha_\tau$ is chosen to minimize over the selected grid the value of the quantile loss wrt to the quantile level, see \cite{gerlach2020recare}.}
In  real applications, this grid search approach can be computationally expensive (depending on the size of the grid) and the performance can be affected by the size and gap of the grid, which is normally decided by means of an ad-hoc approach.

\vspace{0.5cm}
{\centering
\section{\normalsize WEIGHTED QUANTILE ESTIMATORS} \label{model_section}
%\label{s:modprop}
}

\subsection{Motivation}

From the main definition of ES provided in (\ref{e:ESdef}), it follows that that $\text{ES}_{t}$ is related to $F_{t}(.)$ by the following integral
\begin{equation}
    ES_{t}=\frac{1}{F_t(Q_{t,\alpha})}\int_{-\infty}^{Q_{t,\alpha}}rdF_{t}(r)=\frac{1}{\alpha}\int_{-\infty}^{Q_{t,\alpha}}rdF_{t}(r),
\label{e:int1}
\end{equation}
%\int_{-\infty}^{Q_{t,\alpha}}rf_{t}(r)dr
%where $f_{t}(r)=\partial F_t(r) / \partial r$ is the conditional Probability Density Function (PDF) of $r_t$.
that, after a simple change of variable, can be rewritten as
\begin{equation}
    ES_{t}=\frac{1}{\alpha}\int_{0}^{\alpha}Q_{t,p}dp.
\label{e:int2}
\end{equation}
The integral in (\ref{e:int2}) can be approximated over a discrete grid by means of standard numerical integration techniques. Namely, given the target quantile level $\alpha$, assume that an equally spaced grid of quantile levels of size $M$ is selected,
\[
\boldsymbol{\alpha}_M=\left[\alpha_1,\alpha_2,\ldots, \alpha_{M} \right],
\]
where, setting $\alpha_0=0$,
\[
\alpha_m=\alpha_{m-1}+\eta,
\]
with $\alpha_{M}=\alpha$, and $\eta=(\alpha_M-\alpha_1)/(M-1)$, for $m=1,\ldots,M$. A simple \emph{rectangular}  rule would then lead to the following approximation
\begin{equation}
ES_{t}\approx \frac{1}{\alpha}\eta
\sum_{i=1}^{M}Q_{t,\alpha_i}=
\frac{1}{M}\sum_{i=1}^{M}Q_{t,\alpha_i}=\sum_{i=1}^{M}w_iQ_{t,\alpha_i}
\label{e:nuint}
\end{equation}
with $w_i=1/M$, for $i=1,\ldots,M$. It is easy to show that, in theory, as $M\to \infty$ the above approximation asymptotically tends to the ``true'' $\text{ES}_{t}$ value. In general, it can be shown \citep[see][]{rabinowitz1984} that many higher order integration rules, such as the \emph{trapezoidal} and \emph{Simpson's} rule, can be represented as weighted averages of the form in (\ref{e:nuint}) where, modulating the choice of the weights $w_i$, one can obtain different integration rules as special cases. For example, the set of weights $(w_1=1/(2M),w_2=1/M,\ldots,w_{M-1}=1/M, w_M=1/(2M))$ would lead to a \emph{trapezoidal} rule \citep[for more details see][page 57, Section 2.1.5]{rabinowitz1984}. It is however worth noting that, in real data applications, data scarcity prevents accurate estimation of VaR for extreme quantile orders, posing constraints on the choice of the minimum grid value $\alpha_1$.

It follows that a correction for this left-tail truncation bias should be considered when designing an estimator for $\text{ES}_{t}$ based on the representation in Eq. (\ref{e:nuint}). Furthermore, referring to an appropriately defined strictly consistent scoring function, the weights could be estimated rather than fixed. This approach would bring some important advantages. First, it would be possible to modulate, in a data-driven fashion, the weights assigned to each tail-quantile in (\ref{e:nuint}) in order to optimally match the tail properties of returns and, eventually, down-weight less accurately estimated extreme quantiles. Second, it would allow to control the left-tail truncation bias. Last, working with estimated weights would, to some extent, reduce the impact of the subjective choice of the lower bound $\alpha_1$.

Next, starting from a set of consistent estimators of $Q_{t,p}$ ($0 < p \leq \alpha)$, these ideas will be elaborated to define a semi-parametric two-step estimation strategy for $\text{ES}_{t}$.
%In order to simplify notation, in the remainder, unless differently specified, the following notational conventions will be adopted: $ES_{t,\alpha}\equiv ES_{t}$ and $Q_{t,\alpha} \equiv Q_t$\, where $\alpha$ denotes the target level for the estimation of VaR and ES.

\subsection{The proposed approach}

\noindent
In this section, we illustrate a novel two-step approach for the semi-parametric estimation of ES. First, at step 1, we obtain semi-parametric estimates of VaR over a pre-defined set of risk levels $\leq \alpha$. Then, at step-2, conditional on the estimates obtained at step 1 and relying on the representation of ES in Eq. (\ref{e:nuint}), an estimate of the conditional ES is obtained as an affine function of the 1st-stage VaR estimates. For these reasons we refer to our approach as the Weighted Quantile (WQ) estimator. Compared to a simple average, working with an affine transformation offers some important advantages. First, this specification allows to easily control for left-tail truncation bias. Second, since the weights are fitted via the optimization of a strictly consistent scoring function for ($Q_t$, $\text{ES}_t$), it is potentially possible to obtain relevant gains in terms of accuracy in the estimation of VaR and ES.
%\blue{\textbf{ Shall we leave or remove this sentence?}, given by a weighted average plus a constant}.

More precisely, the conditional ES at time $t$ is modelled as
\begin{equation}
ES^{(\text{WQ-Beta})}_{t}= w_{0}+\sum_{i=1}^{M} w_{i} Q_{t,\alpha_i} \, ,
\label{e:insES}
\end{equation}
where the weights $w_{i}$, $i=1,\dots,M$, are generated by some flexible and parsimonious function. Inspired by the literature on mixed data sampling (MIDAS) and distributed lag models \citep[see][among others]{ghysels2007midas}, a suitable choice is given by a Beta weight function\footnote{Compared to the Beta weight functions used in the literature on MIDAS and distributed lag models, a distinctive feature of the specification that is being used in this paper is that weights have not been normalized to sum up to unity.}. Namely, for $i=1,\ldots,M$, we have $w_i=w\left(\frac{i}{M};a,b\right)$ with
\begin{equation}
w(x;a,b)=\frac{x^{a-1}(1-x)^{b-1}\Gamma(a+b)}{\Gamma(a)\Gamma(b)}.
\label{e:betalag}
\end{equation}
The main reasons for adopting the Beta specification to model the weights behaviour in Eq. (\ref{e:insES}) are its parsimony, since it only depends on two parameters, and flexibility. In Appendix \ref{beta_function_section},
Fig. \ref{f:betapatt} displays various patterns that can be generated from the weight structure defined in Eq. (\ref{e:betalag}) for different values of the coefficients $a$ and $b$. Constraining $a$ or $b$ to be equal to 1, a zero-modal behaviour is observed. Namely, for $a=1$ and $b>1$, the Beta weight function returns declining weights while, choosing $a>1$ and $b=1$, increasing weights are obtained. The value of the unconstrained coefficient determines the speed of decay of the curve. Removing the unity constraint on either $a$ or $b$ makes the curve more flexible and allows to reproduce uni-modal, hump-shaped behaviors such as those observed in the lower panel of Fig. \ref{f:betapatt}. Mixtures of Beta polynomials could also be used to further increase the flexibility of the curve, as explored in \cite{ghysels2007midas}.

\subsection{Variants of the weighted quantile ES estimator}\label{all_estimators}
In our empirical investigations, as robustness checks, we also consider some alternative variants of the WQ-Beta framework.
First, we compare the performance of the weighted quantile framework in Eq. (\ref{e:insES}) with a simpler approach based on an equally weighted average of quantiles:
\begin{equation}
ES^{(\text{SA-BC})}_{t}= w_{0}+ \frac{ \sum_{i=1}^{M} Q_{t,\alpha_i} } {M} \, .
\label{e:avgES}
\end{equation}
Here SA-BC represents Simple Average with Bias Correction (using the $w_0$ term).

On the other hand, we also consider a generalization of the WQ estimator, denoted as $\text{ES}^{(\text{WQ-UNC})}_{t}$, in which the $w_i$ weights (in Eq. (\ref{e:insES})) are estimated as free unconstrained (except non-negativity) parameters and not ``indirectly'' recovered from the fitted Beta weight function. Compared to the $\text{ES}^{(\text{WQ-Beta})}_{t}$ approach, which has fixed number of parameters with different $M$, under $\text{ES}^{(\text{WQ-UNC})}_{t}$ the number of fitted $w_i$ parameters is changing with $M$. For example, when $M=10$, $\text{ES}^{(\text{WQ-UNC})}_{t}$ has 10 $w_i$ parameters to be estimated.

Finally, an intermediate step between the above variants of the WQ estimator is to consider the following \emph{equally weighted} specification:
\begin{equation}
ES^{(\text{WQ-EW})}_{t}= w_{0}+ w_{1} \sum_{i=1}^{M} Q_{t,\alpha_i}  \, ,
\label{e:avgESunc}
\end{equation}
that is more general than the $\text{ES}^{(\text{SA-BC})}_{t}$ estimator, since $w_1$ is estimated and the sum of weights ($M w_1$) is not constrained to be equal to 1.

Lastly, to validate the importance of the bias correction (intercept) term $w_0$, we have also tested the following approach based on an equally weighted average of quantiles without the bias correction term:
\begin{equation}
ES^{(\text{SA-No-BC})}_{t}= \frac{ \sum_{i=1}^{M} Q_{t,\alpha_i} } {M} \, .
\label{e:avgES}
\end{equation}

To summarize, there are five different ES estimators to be tested: $\text{ES}^{(\text{WQ-Beta})}_{t}$, $\text{ES}^{(\text{WQ-EW})}_{t}$, $\text{ES}^{(\text{WQ-UNC})}_{t}$, $\text{ES}^{(\text{SA-BC})}_{t}$ and $\text{ES}^{(\text{SA-No-BC})}_{t}$. Their performances will be compared through comprehensive simulation  (Section \ref{simulation_section}) and empirical (Section \ref{data_empirical_section}) studies.

\subsection{Two-step estimation procedure}\label{two_step_estimation}

Next, we provide a detailed description of the two-steps of the proposed ES estimation procedure. Although, for ease of explanation, we focus on the standard risk level $\alpha=2.5\%$, the method can be immediately extended to other values of $\alpha$.\\

\noindent \textbf{Step-1}:

Under step 1, given the target quantile level $\alpha= 2.5\%$, an equally spaced grid of quantile levels of size $M$ is selected,
\[
\boldsymbol{\alpha}_M=\left[\alpha_1,\alpha_2,\ldots, \alpha_{M} \right],
\]
where $\alpha_m=\alpha_{m-1}+\eta$, with  $\alpha_M=\alpha$ and $\eta=(\alpha_M-\alpha_1)/(M-1)$, for $m=2,\ldots,M$. The value of the lower bound $\alpha_1$ can be selected on a case-by-case basis, by taking into account the length of the available in-sample returns series. As an example, with $M=10$ and $\alpha=2.5\%$, fixing $\alpha_1=0.005$ we have $\eta=0.0022$ and the following grid of quantile levels
\[
\boldsymbol{\alpha}_M= \,[0.005, 0.0072, 0.0094, 0.0117, 0.0139, 0.0161, 0.0183,  0.0206, 0.0228, 0.025].
\]
Then, $M$ CAViaR models are separately estimated for each of the quantile orders in $\boldsymbol{\alpha}_M$. In order to overcome the potential quantile crossing problem, the monotonization method proposed by \citet{chernozhukovetal2010} is employed.

In the simulation section, comprehensive studies are conducted to evaluate the impact of $M$ and $\alpha_1$ on the model performance and provide useful guidance on the selection of their values.

For illustrative purposes, without implying any loss of generality, we here refer to the CAViaR symmetric absolute value (CAViaR-SAV) framework
\begin{equation} \label{caviar_model}
Q_{t}= \beta_0+ \beta_1 |r_{t-1}| + \beta_2 Q_{t-1}.
\end{equation}
The proposed procedure can however be immediately extended to consider different conditional quantile models of different nature and complexity, such as CAViaR with asymmetric specification (CAViaR-AS) or nonlinear threshold specifications.

For each trial quantile level $\alpha_m \in \boldsymbol{\alpha_M}$, the above CAViaR-SAV model is then used to produce the time series of conditional in-sample quantiles $\mathbf{Q}_{1:N, \alpha_m}$ and quantile forecasts $\hat{\mathbf{Q}}_{N+1, \alpha_m}$, for $m=1,\ldots,M$. The set of in-sample quantiles at all trial quantile levels, from $\alpha_1$ to $\alpha_M$, is collected in the ${N \times M}$ matrix $\mathbf{Q}_{1:N}$.  %$\mathbf{Q} \in \mathbf{R}^{N \times M}$.
Here, the last ($M$-th) column of $\mathbf{Q}_{1:N}$ corresponds to the time series of 2.5\% in-sample conditional quantiles: $\mathbf{Q}_{1:N, 0.025}$. Similarly, we use the notation $\mathbf{\hat{Q}}_{N+1}$ to indicate the ${1 \times M}$ vector of 1-step-ahead VaR forecasts at all trial quantile levels. The last element in $\mathbf{\hat{Q}}_{N+1}$  represents the VaR forecast at the target $2.5\%$ level: $\hat{Q}_{N+1, 0.025}$.

The parameters of the CAViaR models at the proposed quantile levels $\boldsymbol{\alpha}_M$ can be estimated using a Quasi Maximum Likelihood (QML) approach, following \cite{caviar}. In particular, the quantile regression equation parameters $\left( \beta_{0,\alpha_m},\beta_{1,\alpha_m},\beta_{2,\alpha_m} \right)$ are separately estimated, for each $\alpha_i \in \boldsymbol{\alpha}_M$, by minimizing the quantile loss function:
\begin{equation}\label{q_loss}
\frac{1}{N} \sum_{t=1}^{N}(\alpha_m-I(r_t<Q_{t,\alpha_m}))(r_t-Q_{t,\alpha_m})  \,\, \qquad m=1,\ldots,M,
\end{equation}
whose negative, as shown in \cite{giacominikomunjer2005} among others, can be interpreted as a quasi-likelihood function.

%where $N$ is the sample size, and $Q_{1},Q_{2},\ldots,Q_{N}$ is a series of quantile forecasts at level $\alpha$ for the observations $r_{1},\ldots,r_{T}$.
As documented by \cite{caviar}, solutions to the optimization of the quantile loss objective function can be heavily dependent on the chosen initial values. To account for this issue, we adopt a multi-start optimization procedure inspired by that suggested in \cite{caviar}. First, multiple (10,000 in our paper) candidate parameter starting vectors are generated from uniform random variables defined on the interval [0,1], leading to multiple and different locally optimal QML estimates. Then the top 2 (out of 10,000) sets of the parameters that produced the highest likelihood function values are used as starting values for another optimization round. This was performed using the interior-point algorithm implemented in the Matlab \emph{fmincon} function\footnote{All computations were performed using Matlab 2019b.}. Lastly, the final parameter estimates are selected as the ones producing higher objective function values from the 2 sets of starting values.

We would like to emphasize that the proposed framework can actually incorporate quantile estimates obtained from any model (not necessarily CAViaR), while we leave this for future research.

%, we have $\mathbf{\hat{Q}} \in \mathbf{R}^{M \times 1}$
%which is $\mathbf{\hat{Q}}(M)$,

\noindent \textbf{Step-2}:

In the second stage of our approach, we predict the conditional ES as an affine function of the elements of $\hat{\mathbf{Q}}_{N+1}$. Focusing, for ease of presentation, on the $\text{ES}^{(\text{WQ-Beta})}$ estimator, the only unknown parameters in Eq. (\ref{e:insES}) are $(w_0,a,b)$. Conditioning on first stage fitted VaR series $\mathbf{Q}_{1:N}$, these can be estimated minimizing the strictly consistent scoring function:
\[
(\hat{w}_0,\hat{a},\hat{b})=\underset{(w_0,a,b)}{\arg\min}{\sum_{t=1}^{N}S_t\left(r_t,
\hat{Q_t}, ES_t(w_0,a,b|\hat{\mathbf{Q}}_t)\right)},
\]
where $S_t(.)$ is defined as in Eq. (\ref{es_caviar_log_score}) and $\text{ES}_t$ follows the specification in Eq. (\ref{e:insES}).

%\begin{eqnarray*}
%S_t(r_t, ES_t;w_0,a,b|\mathbf{Q}_t) &=& (I_t -\alpha)G_1(Q_{t}) - I_tG_1(r_t) +  G_2(ES_t)\left(ES_t-Q_t + \frac{I_t}{\alpha}(Q_t-r_t)\right)
%\nonumber
%\\
%                      &-& H(ES_t) + a(r_t) \, ,
%                      \label{e:estscore}
%\end{eqnarray*}
Minimization of the above AL log-score was implemented using the Quasi-Newton optimizer implemented in the Matlab \emph{fminunc} function. The number of grid points $M$ and the lower bound $\alpha_1$ are hyper-parameters that we need to choose. Overall, $M$ affects the trade-off between accuracy and computational cost while the choice of the lower bound $\alpha_1$ could eventually affect the bias properties of the estimated ES. In Section \ref{simulation_section} comprehensive simulation studies are conducted for assessing the effects of incorporating various choices of $M$ and $\alpha_1$. However, our weighted quantile framework is turned out to be capable of accurately predicting the ES using a very small value of $M$, i.e. $M=3$. Furthermore, the simulation suggests that the bias of the proposed WQ estimator does not crucially depend on the choice of $\alpha_1$.

One-step-ahead forecasts of $\text{ES}_t$ can then be easily computed by replacing estimated in-sample quantiles, on the RHS of (\ref{e:insES}), by their out-of-sample forecasts obtained from the associated CAViaR models ($\mathbf{\hat{Q}}_{N+1}$). Formally, the ES predictor at time $N+1$, conditional on in-sample information available at time $N$, is obtained as
\begin{equation}
\widehat{ES}_{N+1,\alpha}= w_{0,N}+\sum_{i=1}^{M} w_{i,N} \hat{Q}_{N+1,\alpha_i} \, ,
\label{e:outsES}
\end{equation}
where the subscript $N$ in $w_{i,N}$ indicates that the weight function is estimated using information up to time $N$. \\

\noindent
\emph{Remark 1.} As previously discussed, the estimation of ES based on numerical integration of the tail quantiles is inherently affected by truncation bias since the summation on the RHS of (\ref{e:insES}) does not involve conditional quantiles of order below $\alpha_1$. In Eq. (\ref{e:insES}) we control for the truncation bias in two different ways. First, we include the intercept term $w_0$ in order to control for \emph{fixed} bias. Second, the sum of weights appearing on the RHS of  (\ref{e:insES}), $\theta=\sum_{i=1}^{M}w_i$, has been deliberately left unconstrained in order to allow the size of bias to depend on the average tail VaR level.
%\red{In Section \ref{simulation_section}, the impact of different bias sources related to truncation, discretization and misspecification, respectively, will be numerically assessed via a Monte Carlo simulation.}
% \begin{flushright}
% $\diamondsuit$
% \end{flushright}

\noindent
\emph{Remark 2.} It is worth noting that, letting $\tilde{w}_i=w_i/\theta$,  Eq. (\ref{e:insES}) can be alternatively written as
\begin{equation}
ES_{t}= w_{0}+\theta\sum_{i=1}^{M} \tilde{w}_{i} Q_{t,\alpha_i} \, ,
\label{e:insES2}
\end{equation}
where $\sum_{i=1}^{M}\tilde{w}_i=1$ by construction. The reparameterization in (\ref{e:insES2}) makes evident the role of $\theta$ for bias correction.
% \begin{flushright}
% $\diamondsuit$
% \end{flushright}
%In the simulation study, to demonstrate the effect of $M$, we have tested $M=3$, $M=5$, $M= 10$ and $M=50$ respectively.

\noindent
\emph{Remark 3.} As presented in Fig. \ref{f:betapatt} in Appendix \ref{beta_function_section}, the last element of the sequence of weights generated from the Beta weight function (i.e. the one corresponding to the risk level $\alpha_M=\alpha$) is by construction equal to 0, except when parameter $b$ equals to 1\footnote{In this case, it is equal to 1 under the convention $0^0=1$. In the practical implementation on both real and simulated data, we find that estimated $b$ is never exactly 1, thus we have the last weight in Beta weight function always as 0.}. To address this issue, in the implementation of WQ-Beta, we set the number of grid points equal to $M+1$, so that the weight of the $\alpha_{M}$-quantile is not 0 by construction. Referring to the previous example, to estimate ES at level $\alpha=2.5\%$ as a weighted average of $M=10$ tail quantiles of order $\alpha_i\leq\alpha$, setting $\alpha_1=0.005$, we simply use an equally spaced grid of $M+1=11$ points, still with $\alpha_M=0.025$.
\[
\boldsymbol{\alpha}_{M+1}= \,[0.005, 0.0072, 0.0094, 0.0117, 0.0139, 0.0161, 0.0183,  0.0206, 0.0228, \mathbf{0.025}, 0.0272].
\]
At no additional cost in terms of estimated parameters, this simple solution guarantees that the estimated weight for quantile at level $\alpha_{M+1}=0.0272$ will  always be 0 while assigning a positive weight to the $\alpha_M=0.025$ quantile. In this way, consistently with its theoretical definition, the ES will be estimated as a weighted average of $M$ quantiles of order $\alpha_i\leq \alpha$ ($i=1,\ldots,M$) without systematically excluding the estimated VaR at the target level $\alpha$.
% \begin{flushright}
% $\diamondsuit$
% \end{flushright}

%\blue{\textbf{I would remove this} Moving the focus on the estimation of the ES, when the weighted quantile estimator $ES^{(wq)}_t$ in (\ref{e:insES}) is considered, the parameters to be estimated are the intercept term $w_{0}$ and the coefficients of the beta function ($a, b$). \blue{Conditional on first stage VaR estimates, these are estimated minimizing the AL log score function defined in (\ref{es_caviar_log_score}) using the Quasi-Newton optimizer implemented in the Matlab \emph{fminunc} function.} Differently, for the bias corrected simple average estimator $ES^{(avg)}_t$ in  (\ref{e:avgES}), the only parameter to be estimated is intercept term $w_{0}$ which is also estimated by unconstrained optimization.}

\noindent
\emph{Remark 4.} The estimated conditional ES series obtained through the Weighted Quantile estimator have simple and easily interpretable dynamic properties.
%Assuming, for ease of presentation, that the first-stage VaR series are generated by a CAViaR-SAV model,
It is easy to show that Eq. (\ref{e:insES}) can be rewritten as
\begin{eqnarray}
ES_{t}= w_{0}+ \bar{\beta}_0 + \bar{\beta}_1|r_{t-1}|+
\sum_{i=1}^{M} w_{i} \beta_{2,i}Q_{t-1,\alpha_i}\,
\label{e:ESdyn1}
\end{eqnarray}
where $\bar{\beta}_k=\sum_{i=1}^{M} w_{i} \beta_{k,i}$, for $k=0,1$, and $(\beta_{0,i},\beta_{1,i}, \beta_{2,i})$  are the parameters of the CAViaR-SAV model for the conditional $\alpha_i$-quantile of $r_t$, for $i=1,\ldots,M$. If the CAViaR-SAV model is well specified, i.e. if log-returns are generated by the following GARCH-type process
\begin{eqnarray*}
r_t &=& \sigma_t \varepsilon_t, \qquad \varepsilon_t \stackrel{\rm i.i.d.}{\sim}(0,1), \\
\sigma_t &=& \omega + \gamma |r_{t-1}| + \delta \sigma_{t-1},\qquad \omega>0\,,\gamma >0,\, \delta >0,
\end{eqnarray*}
$\beta_{2,i}$ will be constant across different quantile orders, i.e. $\beta_{2,i}=\bar{\beta}_{2}=\delta$, for $i=1,\ldots,M$, as also largely confirmed by our empirical results on real financial data.

%Equation (\ref{e:ESdyn1}) can then be rewritten as
%\begin{eqnarray}
%ES_{t}= w_{0}+ \bar{\beta}_0 + %\bar{\beta}_1|r_{t-1}|+  \bar{\beta_{2}}ES_{t-1}\, %\label{e:ESdyn2}
%\end{eqnarray}

%If we introduce the additional constraint $\sum_{i=1}^{M}w_1 = 1$, we have $\min{\beta_{k,i}}\leq \bar{\beta}_k \leq \max{\beta_{k,i}}$, for $i=0,1$, $i=1,\ldots,M$, by construction.
%On the other hand, if $\sum_{i=1}^{M}w_1>1$
%If we assume $\beta_{2,i}$ to be constant across different quantile orders ($\beta_{2,i}=\bar{\beta}_{2}$), as largely confirmed by our empirical results on real financial data,
Eq. (\ref{e:ESdyn1}) will then simplify to the following
\begin{eqnarray}
ES_{t} &=& w_{0}+ \bar{\beta}_0 + \bar{\beta}_1|r_{t-1}|+
\bar{\beta}_{2}\sum_{i=1}^{M} w_{i} Q_{t-1,\alpha_i}\, \\ \nonumber
&=& w_{0}+ \bar{\beta}_0 + \bar{\beta}_1|r_{t-1}|+
\bar{\beta}_{2}(ES_{t-1}-w_0)\\ \nonumber
&=& \beta^*_0 + \bar{\beta}_1|r_{t-1}|+
\bar{\beta}_{2}ES_{t-1}
\label{e:ESdyn2}
\end{eqnarray}
where $\beta^*_0=w_{0}(1-\bar{\beta}_2) + \bar{\beta}_0$. These derivations show that, in our approach, the ES is allowed to have dynamics that are separate from those of VaR. At the same time, these are automatically implied by the dynamics of conditional quantiles in the tail below VaR, without requiring any additional ad-hoc assumptions.
% \begin{flushright}
% $\diamondsuit$
% \end{flushright}

%In the empirical section, we found that this framework is however consistently outperformed by the more complex weighted quantile estimator.

%\bigskip
%{\centering
%\section{\normalsize Implied ES Dynamics} \label{s:dynamo}
%\par
%}
%\noindent

%\textbf{ Also, as shown in Section \ref{s:modprop}, well-definiteness of the estimated $(VaR,ES)$ can be easily ensured by means of simple, theoretically motivated, parametric constraints.}

Comparing our approach to other existing proposals, it should be remarked that our weighted quantile estimator is more flexible than that proposed in Model (\ref{es_caviar_mult_model}) by \cite{tayl2017}, based on the assumption that the conditional ES is a multiplicative rescaling of the fitted VaR model. Also, it differs from the ``additive'' approach proposed in Model (\ref{es_caviar_add_model}) of the same paper under two main respects. First, we directly model the dynamics of the ES rather than the difference between ES and VaR. Second, the ES estimates are continuously updated, and not only when VaR is violated.

\bigskip
{\centering
\section{\normalsize SIMULATION} \label{simulation_section}
\par
}
\noindent

In this section, simulation studies are conducted to assess the statistical properties and performances of the proposed models, with respect to the one-step-ahead $\alpha= 2.5\%$ VaR and ES forecast accuracy.

In the weighted quantile approach, the ES estimation and forecast bias is potentially related to three factors: (1) truncation which is related to the hyper-parameter $\alpha_1$; (2) discretization which is related to hyper-parameter $M$; (3) misspecified conditional quantile dynamics. Regarding (3), since CAViaR-SAV is used as the quantile estimation approach, the corresponding ``true'' GARCH model is the Absolute Value (AV) GARCH. Thus, if a different GARCH type model is used as the Data Generating Process (DGP), we can evaluate the impact of misspecified conditional quantile dynamics on the VaR and ES estimation and forecast.

Therefore, to comprehensively evaluate the performance of the proposed framework, in the simulation study we have tested various implementation settings with different $\alpha_1$, $M$ and DGPs.

%Namely, to compare the bias and efficiency of the proposed weighted and simple average quantile methods, both the mean and Root Mean Squared Error (RMSE) values are calculated over the replicated time series.

The simulation design is structured as follows: 1000 replicated return series are generated from a AV-GARCH-t and a GARCH-t. The model specifications are presented as below. \\

\noindent
\textbf{Simulation Model 1: (AV GARCH-t)}
\begin{eqnarray} \label{av_garch_simu}
r_t&=& \sigma_t \varepsilon_t, \\ \nonumber
\sigma_t&=& 0.02 + 0.10 |r_{t-1}| + 0.85 \sigma_t, \\\nonumber
\end{eqnarray}

\noindent
\textbf{Simulation Model 2: (GARCH-t)}
\begin{eqnarray} \label{garch_simu}
r_t&=& \sigma_t \varepsilon_t, \\ \nonumber
\sigma_t^2&=& 0.02 + 0.10 r_{t-1}^2 + 0.85 \sigma_t^2,
\end{eqnarray}

For both models, $\varepsilon_t \stackrel{\rm i.i.d.} {\sim} t_\nu(0,1)$ with $\nu$ indicating the DoF parameter. The value of the DoF parameter has been chosen be 10 which aims to match, on average, its real data estimates in our empirical study. The other parameters have also been set equal to values close to their empirical estimates.

%\textbf{This is not clear to me yet: $n=1900$ is approximately the average in-sample (fixed) size for the empirical study across 7 indices, details as in Table ...... . To match up with the forecasting study and to find properties for the estimators in a similar situation, $n=1900$ is selected as the sample size in the simulation study.

In the same spirit, to facilitate the comparisons with the findings of our empirical study, the simulation has been performed considering as sample size $n=1900$, that has been chosen to approximately match the length of the available in-sample period in our empirical application in Section \ref{data_empirical_section}.

%\textbf{Probably we could only use one 1900 sample size, as we already have 8 tables and the VaR and ES forecasts that are very close to true. Then we might leave the asymptotic discussion as future work. And it might be a good idea to merge the different tables using different lags, to make the presentation clearer.}

%\textbf{The simulation results with models using different lags are now merged.}

The \emph{true} one-step-ahead level $\alpha$ VaR forecasts from the above simulation models are calculated as:
\begin{equation}\label{var_t_dist_1}
Q_{t+1}=  \sigma_{t+1} t_{\nu}^{-1}(\alpha)\sqrt{\frac{\nu-2}{\nu}}, \nonumber
\end{equation}
where $t_{\nu}^{-1}$ is the inverse of Student-t's CDF with the $\nu$ degrees of freedom. ES forecasts from the same model are calculated as:

\begin{equation}\label{es_t_dist_1}
\text{ES}_{t+1}=  -\sigma_{t+1}  \left( \frac{g_{\nu}(t_{\nu}^{-1}(\alpha))}{\alpha} \right) \left( \frac{\nu+ (t_{\nu}^{-1}(\alpha))^{2}}{\nu-1} \right) \sqrt{\frac{\nu-2}{\nu}}, \nonumber
\end{equation}
where $g_{\nu}$ is the Student-t PDF.

These \emph{true} VaR and ES forecasts are calculated for each data set and used to compute bias for the VaR and ES forecasts obtained from a CAViaR-SAV, for the VaR, and both the $\text{ES}^{(\text{WQ-Beta})}$ estimator and its variants presented in Section \ref{all_estimators}, for the ES. The averages of the true and estimated VaR and ES forecasts, over the 1000 time series, are calculated for both DGPs. More specifically, for AV-GARCH-t, we have \emph{true} $Q_{t+1}= -1.3775$ and \emph{true} $\text{ES}_{t+1}= -1.7428$. For GARCH-t, we have \emph{true} $Q_{t+1}= -1.9079$ and \emph{true} $\text{ES}_{t+1}= -2.4138$.

Five different models are tested: $\text{ES}^{(\text{WQ-Beta})}_{t}$, $\text{ES}^{(\text{WQ-UNC})}_{t}$,  $\text{ES}^{(\text{WQ-EW})}_{t}$, $\text{ES}^{(\text{SA-BC})}_{t}$ and $\text{ES}^{(\text{SA-No-BC})}_{t}$, with $M\in\{3,5,10,50\}$. Therefore, to remark the dependence on the value of $M$, we use the notation  WQ-Beta-M, WQ-EW-M and WQ-UNC-M, SA-BC-M and SA-No-BC-M to indicate estimators based on a grid size equal to $M$. We have also tested $\alpha_1\in\{0.005,0.01,0.015\}$. In addition, we use $\text{VaR}_\Delta$ and $\text{ES}_\Delta$ to represent the mean absolute deviations between VaR and ES \emph{true} and estimated forecasts, respectively.

Regarding the quantile forecast which is based on the CAViaR-SAV model, the $\text{VaR}_{\Delta}$ is in general quite small which lends evidence on the Step-1 quantile estimation process as described in Section \ref{two_step_estimation}. More specifically, for Simulation Model 1 (AV-GARCH-t), the $\text{VaR}_{\Delta}$ is 0.0019. With respect to Simulation Model 2 (GARCH-t), the $\text{VaR}_{\Delta}$ is 0.0052 which is larger that of Simulation Model 1 and as expected, since we choose the GARCH-t as the DGP to evaluate the impact of misspecified conditional quantile dynamics on the VaR and ES estimation and forecast.

The ES forecasting simulation results are summarized in Table \ref{simu_table_av} and \ref{simu_table_garch}, for each model, each $\alpha_1$ and each $M$. A box highlights the smallest $\text{ES}_\Delta$ for each row.

Focusing on the ES forecasting results in Tables \ref{simu_table_av} and  \ref{simu_table_garch}, we first note that the weighted quantile estimators produce quite small $\text{ES}_\Delta$, which supports their validity for ES estimation and forecasting. Second, we observe that the weighted quantile approaches produce smaller $\text{ES}_\Delta$ than the simple average approaches. Third, among the weighted quantile estimators, the more flexible WQ-Beta and WQ-UNC produce better accuracy than WQ-EW. Lastly, comparing WQ-Beta and WQ-UNC, for different $M$ and $\alpha_1$, in general WQ-Beta produces close or better $\text{ES}_\Delta$ results than WQ-UNC. This lends support on the usefulness of incorporating the Beta weight function as the weighting scheme.

The simulation results clearly favor the weighted quantile estimator, compared to the simple average approaches SA-BC and SA-No-BC. The SA-No-BC approach clearly has the largest $\text{ES}_\Delta$, especially when $\alpha_1$ is larger, which informs the importance of the bias correction process. In addition, since the weighted quantile approaches in general outperform the SA-BC approach, thus we can see the additional gain by incorporating the weighted quantiles.

Regarding the impact of $M$, which is the number of averaged quantiles, it is interesting to note that its choice is not critical and, in general, the weighted quantile estimators based on only $M=3$ grid points (WQ-Beta-3) are already characterized by good performances. In addition, using $M=5, 10 \text{ and } 50$, we can still observe accurate and very close ES forecasting results. However, the $M=50$ setting requires much higher computational cost compared to other options. Therefore, we have selected $M=3, 5 \text{ and } 10$ for the empirical study.

Regarding the impact of $\alpha_1$, which is the lower bound of the chosen grid of quantile levels, we overall observe that our weighted quantile models produce accurate ES forecasts for all three choices of $\alpha_1$. This again lends evidence on the effectiveness on the weighted quantile (and the included bias correction) process. In other words, WQ methods have a limited dependence on the lower bound, while this does not happen for the simple average models. As highlighted in Tables \ref{simu_table_av} and \ref{simu_table_garch}, in general the choice $\alpha_1=0.005$, which is the one that is closest to 0, produces the smallest $\text{ES}_\Delta$ for different approaches. This result, consistent with the ES integral definition in Eq. (\ref{e:int2}) which has the integral lower bound as zero, drives the choice of $\alpha_1=0.005$ for our empirical study.

%Such results means that in the empirical study the choice of $\alpha_1$ is relatively straightward, i.e., choosing an $\alpha_1$ that is close to 0.

% these tend to further improve as the number of averaged quantiles M increases. However, gains tend to decrease as the value of M increases. In particular, the simulation results show that increasing the value of M from 10 to 50 has minimal impact on estimation accuracy while still leading to a remarkable increase of computational time.

\begin{figure}[htp]
	\centering
	\includegraphics[width=0.9\textwidth]{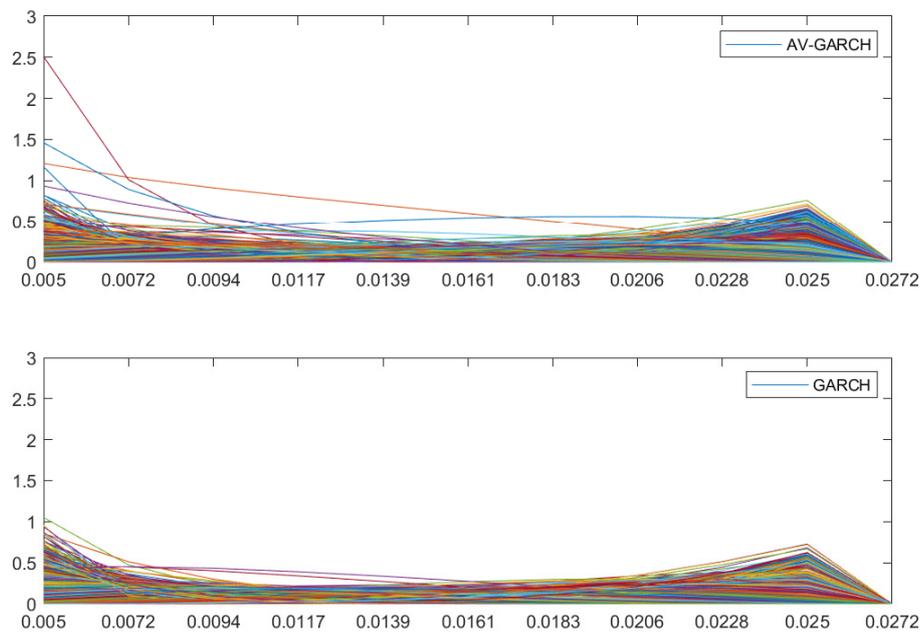}
	\caption{\label{Fig_beta_simu}Estimated Beta weights for the 1000 simulated time series of each data generating process, $M=10$.}
\end{figure}

The estimated Beta weights for the 1000 simulated time series of each DGP are presented in Fig. \ref{Fig_beta_simu}. The fitted weights distribution is characterized by two modes respectively occurring at the lower truncation point of the selected grid of $\alpha$ values and immediately before the upper truncation point, that is the target ES order. The lower mode is evidently accounting for the truncation bias arising from the omission of extreme left quantiles. Furthermore, in the absence of left truncation bias, we conjecture the pattern of the Beta weights would be expected to match the profile of the returns tail distribution. The simulation results also lend potential support on this conjecture.

%Table \ref{simu_table_av} summarizes the simulated distribution of the estimated coefficients for the different settings of the $ES^{(\text{WQ-Beta})}$ and $ES^{(\text{SA-BC})}$ estimators that have been here considered. For SA-BC estimators, the average estimated ${w}_0$ intercept is, as expected, negative, correcting for the left tail truncation bias. Confirming our intuition, this is more substantial for heavy tailed processes that are for low values of $\nu$.

%WQ estimators are more flexible since the bias correction takes place through both ${w}_0$ and $\theta=\sum_{i=1}^{M}w_i$. The average value of the estimated intercept is positive being compensated by the fact that the average estimated $\theta$ is greater than 1, as expected. Again, in line with our findings for the SA-BC estimator, the difference $(\theta-1)$ is higher for lower values of the degrees of freedom parameter $\nu$.

Overall, the simulation results illustrate the validity of the proposed weighted quantile models and the corresponding estimation process. The performance of the weighted and simple average quantile approaches will be further compared in the empirical section.

\begin{table}[!ht]
\begin{center}
\caption{\label{simu_table_av} \small ES forecast results with various $\alpha_1$ and $M$ from proposed models, with data simulated from Simulation Model 1 (AV GARCH-t). True $\text{ES}_{t+1}= -1.7428$.}
\tabcolsep=10pt
\small
\begin{tabular}{lcccccc} \hline
ES forecast $\Delta$& $\alpha_1= 0.005$  & $\alpha_1=0.01$ & $\alpha_1=0.015$ &\\\hline
% $\text{VaR}_{t+1}$&-1.3791&-1.3802&-1.379&\\
% WQ-Beta-3 $\text{ES}_{t+1}$&-1.7591&-1.7498&-1.7421&\\
% SA-BC-3 $\text{ES}_{t+1}$&-1.7284&-1.7244&-1.7219&\\
% SA-No-BC-3 $\text{ES}_{t+1}$&-1.6372&-1.5339&-1.4692&\\
% WQ-Equal-3 $\text{ES}_{t+1}$&-1.7343&-1.7346&-1.735&\\
% WQ-UNC-3 $\text{ES}_{t+1}$&-1.736&-1.7357&-1.7351&\\
% $\text{VaR}_{\Delta}$&0.0016 &0.0027& \fbox{0.0015}&\\
WQ-Beta-3 $\text{ES}_\Delta$&0.0163&0.0070& \fbox{0.0007}&\\
WQ-EW-3 $\text{ES}_\Delta$&0.0085&0.0082&\fbox{0.0078}&\\
WQ-UNC-3 $\text{ES}_\Delta$&\fbox{0.0068}&0.0071&0.0076&\\
SA-BC-3 $\text{ES}_\Delta$&\fbox{0.0144}&0.0184&0.0208&\\
SA-No-BC-3 $\text{ES}_\Delta$&\fbox{0.1055}&0.2088&0.2736&\\\hline
% $\Delta$ Average&0.0303&0.0499&0.0621&\\ \hline
% $\text{VaR}_{t+1}$&-1.3783&-1.3793&-1.3795&\\
% WQ-Beta-5 $\text{ES}_{t+1}$&-1.7431&-1.74&-1.7371&\\
% SA-BC-5 $\text{ES}_{t+1}$&-1.728&-1.7245&-1.7219&\\
% SA-No-BC-5 $\text{ES}_{t+1}$&-1.6174&-1.5282&-1.4672&\\
% WQ-Equal-5 $\text{ES}_{t+1}$&-1.7345&-1.7349&-1.7346&\\
% WQ-UNC-5 $\text{ES}_{t+1}$&-1.7362&-1.736&-1.7349&\\
%$\text{VaR}_{\Delta}$&\fbox{0.0008}&0.0018&0.0020&\\
WQ-Beta-5 $\text{ES}_\Delta$&\fbox{0.0003}&0.0028&0.0057&\\
WQ-EW-5 $\text{ES}_\Delta$&0.0083&\fbox{0.0079}&0.0082&\\
WQ-UNC-5 $\text{ES}_\Delta$&\fbox{0.0066}&0.0067&0.0079&\\
SA-BC-5 $\text{ES}_\Delta$&\fbox{0.0148}&0.0183&0.0209&\\
SA-No-BC-5 $\text{ES}_\Delta$&\fbox{0.1254}&0.2146&0.2756&\\\hline
% $\Delta$ Average&0.0311&0.0501&0.0637&\\ \hline
% $\text{VaR}_{t+1}$&-1.3794&-1.3787&-1.3808&\\
% WQ-Beta-10 $\text{ES}_{t+1}$&-1.7378&-1.7359&-1.7356&\\
% SA-BC-10 $\text{ES}_{t+1}$&-1.7267&-1.7235&-1.7218&\\
% SA-No-BC-10 $\text{ES}_{t+1}$&-1.6054&-1.5239&-1.4662&\\
% WQ-Equal-10 $\text{ES}_{t+1}$&-1.7343&-1.7342&-1.7346&\\
% WQ-UNC-10 $\text{ES}_{t+1}$&-1.7362&-1.7354&-1.7353&\\
%$\text{VaR}_{\Delta}$&0.0019&\fbox{0.0012}&0.0033&\\
WQ-Beta-10 $\text{ES}_\Delta$&\fbox{0.0050}&0.0069&0.0072&\\
WQ-EW-10 $\text{ES}_\Delta$&0.0085&0.0086&\fbox{0.0082}&\\
WQ-UNC-10 $\text{ES}_\Delta$&\fbox{0.0066}&0.0074&0.0075&\\
SA-BC-10 $\text{ES}_\Delta$&\fbox{0.0161}&0.0193&0.0209&\\
SA-No-BC-10 $\text{ES}_\Delta$&\fbox{0.1373}&0.2189&0.2766&\\ \hline
% $\Delta$ Average&0.0347&0.0522&0.0641&\\ \hline
% $\text{VaR}_{t+1}$&-1.3791&-1.3798&-1.3791&\\
% WQ-Beta-50 $\text{ES}_{t+1}$&-1.7356&-1.7357&-1.7351&\\
% SA-BC-50 $\text{ES}_{t+1}$&-1.7269&-1.7238&-1.7216&\\
% SA-No-BC-50 $\text{ES}_{t+1}$&-1.5984&-1.5219&-1.4653&\\
% WQ-Equal-50 $\text{ES}_{t+1}$&-1.7344&-1.7345&-1.7343&\\
% WQ-UNC-50 $\text{ES}_{t+1}$&-1.7361&-1.736&-1.7352&\\
%$\text{VaR}_{\Delta}$&0.0016&0.0023&0.0016&\\
WQ-Beta-50 $\text{ES}_\Delta$&0.0072&\fbox{0.0071}&0.0077&\\
WQ-EW-50 $\text{ES}_\Delta$&0.0084&\fbox{0.0083}&0.0085&\\
WQ-UNC-50 $\text{ES}_\Delta$&\fbox{0.0067}&0.0068&0.0075&\\
SA-BC-50 $\text{ES}_\Delta$&\fbox{0.0159}&0.0190&0.0212&\\
SA-No-BC-50 $\text{ES}_\Delta$&\fbox{0.1444}&0.2209&0.2775&\\\hline
% $\Delta$ Average&0.0365&0.0524&0.0645&\\ \hline
\end{tabular}
\end{center}
\footnotesize \emph{Note}: A box highlights the smallest $\text{ES}_\Delta$ for each row. Note that for WQ-Beta models the estimation grid actually includes $M+1$ values, that is $M+1= 4, 6, 11 \text{ and } 51$, with the last Beta weight being equal to 0 by construction, see Remark 3 for details.
\end{table}

\begin{table}[!ht]
\begin{center}
\caption{\label{simu_table_garch} \small ES forecast results with various $\alpha_1$ and $M$ from proposed models, with data simulated from Simulation Model 2 (GARCH-t). True $\text{ES}_{t+1}= -2.4138$.}
\tabcolsep=10pt
\small
\begin{tabular}{lcccccc} \hline
ES forecast $\Delta$& $\alpha_1= 0.005$  & $\alpha_1=0.01$ & $\alpha_1=0.015$ &\\\hline
% $\text{VaR}_{t+1}$&-1.9026&-1.9022&-1.9027&\\
% WQ-Beta-3 $\text{ES}_{t+1}$&-2.4238&-2.41&-2.4&\\
% SA-BC-3 $\text{ES}_{t+1}$&-2.3865&-2.3794&-2.3774&\\
% SA-No-BC-3 $\text{ES}_{t+1}$&-2.2561&-2.1112&-2.0243&\\
% WQ-Equal-3 $\text{ES}_{t+1}$&-2.3892&-2.3894&-2.3908&\\
% WQ-UNC-3 $\text{ES}_{t+1}$&-2.3917&-2.3908&-2.3905&\\
%$\text{VaR}_{\Delta}$&0.0053&0.0057&0.0052&\\
WQ-Beta-3 $\text{ES}_\Delta$&\fbox{0.0100}&0.0038&0.0138&\\
WQ-EW-3 $\text{ES}_\Delta$&0.0245&0.0244&\fbox{0.0229}&\\
WQ-UNC-3 $\text{ES}_\Delta$&\fbox{0.0221}&0.0230&0.0233&\\
SA-BC-3 $\text{ES}_\Delta$&\fbox{0.0272}&0.0344&0.0364&\\
SA-No-BC-3 $\text{ES}_\Delta$&\fbox{0.1577}&0.3026&0.3895&\\\hline
% $\Delta$ Average&0.0483&0.0776&0.0972&\\\hline
% $\text{VaR}_{t+1}$&-1.9024&-1.903&-1.9024&\\
% WQ-Beta-5 $\text{ES}_{t+1}$&-2.4001&-2.3973&-2.3937&\\
% SA-BC-5 $\text{ES}_{t+1}$&-2.384&-2.38&-2.3773&\\
% SA-No-BC-5 $\text{ES}_{t+1}$&-2.2264&-2.1041&-2.0214&\\
% WQ-Equal-5 $\text{ES}_{t+1}$&-2.3887&-2.3897&-2.3908&\\
% WQ-UNC-5 $\text{ES}_{t+1}$&-2.3922&-2.3915&-2.391&\\
%$\text{VaR}_{\Delta}$&0.0054&0.0048&0.0054&\\
WQ-Beta-5 $\text{ES}_\Delta$&\fbox{0.0137}&0.0165&0.0201&\\
WQ-EW-5 $\text{ES}_\Delta$&0.0251&0.0241&\fbox{0.0230}&\\
WQ-UNC-5 $\text{ES}_\Delta$&\fbox{0.0216}&0.0223&0.0227&\\
SA-BC-5 $\text{ES}_\Delta$&\fbox{0.0298}&0.0338&0.0365&\\
SA-No-BC-5 $\text{ES}_\Delta$&\fbox{0.1873}&0.3097&0.3923&\\\hline
% $\Delta$ Average&0.0555&0.0813&0.0989&\\\hline
% $\text{VaR}_{t+1}$&-1.9027&-1.9026&-1.9028&\\
% WQ-Beta-10 $\text{ES}_{t+1}$&-2.394&-2.3928&-2.3921&\\
% SA-BC-10 $\text{ES}_{t+1}$&-2.3844&-2.3801&-2.3772&\\
% SA-No-BC-10 $\text{ES}_{t+1}$&-2.2124&-2.0999&-2.0201&\\
% WQ-Equal-10 $\text{ES}_{t+1}$&-2.3892&-2.3902&-2.3907&\\
% WQ-UNC-10 $\text{ES}_{t+1}$&-2.392&-2.3915&-2.3911&\\
%$\text{VaR}_{\Delta}$&0.0052&0.0052&0.0051&\\
WQ-Beta-10 $\text{ES}_\Delta$&\fbox{0.0198}&0.0210&0.0217&\\
WQ-EW-10 $\text{ES}_\Delta$&0.0246&0.0236&\fbox{0.0231}&\\
WQ-UNC-10 $\text{ES}_\Delta$&\fbox{0.0218}&0.0223&0.0227&\\
SA-BC-10 $\text{ES}_\Delta$&\fbox{0.0294}&0.0337&0.0366&\\
SA-No-BC-10 $\text{ES}_\Delta$&\fbox{0.2014}&0.3139&0.3937&\\\hline
% $\Delta$ Average&0.0594&0.0829&0.0996&\\\hline
% $\text{VaR}_{t+1}$&-1.9027&-1.9028&-1.9026&\\
% WQ-Beta-50 $\text{ES}_{t+1}$&-2.3912&-2.3912&-2.3915&\\
% SA-BC-50 $\text{ES}_{t+1}$&-2.3842&-2.38&-2.377&\\
% SA-No-BC-50 $\text{ES}_{t+1}$&-2.2025&-2.0967&-2.0189&\\
% WQ-Equal-50 $\text{ES}_{t+1}$&-2.3892&-2.3899&-2.3904&\\
% WQ-UNC-50 $\text{ES}_{t+1}$&-2.3925&-2.392&-2.3912&\\
%$\text{VaR}_{\Delta}$&0.0052&0.0051&0.0053&\\
WQ-Beta-50 $\text{ES}_\Delta$&0.0226&0.0226&\fbox{0.0222}&\\
WQ-EW-50 $\text{ES}_\Delta$&0.0246&0.0238&\fbox{0.0234}&\\
WQ-UNC-50 $\text{ES}_\Delta$&\fbox{0.0213}&0.0218&0.0226&\\
SA-BC-50 $\text{ES}_\Delta$&\fbox{0.0296}&0.0337&0.0368&\\
SA-No-BC-50 $\text{ES}_\Delta$&\fbox{0.2113}&0.3171&0.3949&\\\hline
% $\Delta$ Average&0.0619&0.0838&0.1&\\\hline
\end{tabular}
\end{center}
\footnotesize \emph{Note}: A box highlights the smallest $\text{ES}_\Delta$ for each row. Note that for WQ-Beta models the estimation grid actually includes $M+1$ values, that is $M+1= 4, 6, 11 \text{ and } 51$, with the last Beta weight being equal to 0 by construction, see Remark 3 for details.
\end{table}

\captionsetup[figure]{labelformat={default},labelsep=period,name={Fig.}}

\clearpage
{\centering
\section{\normalsize EMPIRICAL STUDY}\label{data_empirical_section}
\par
}

\subsection{Data and empirical study design}\label{data_section}
The daily data, including open, high, low and closing prices, are downloaded from Thomson Reuters Tick History and cover the period from the beginning of 2000 to the end of 2015. Data are collected for 7 market indices: S\&P500, NASDAQ (both US), Hang Seng (Hong Kong), FTSE 100 (UK), DAX (Germany), SMI (Swiss) and ASX200 (Australia).
%The starting date of the data is chosen as early as 2000.
%$\alpha= 2.5\%$ is employed for both one day ahead VaR and ES forecasting study for the 7 indices.

A rolling window with fixed in-sample size is employed for estimation to produce each one-step-ahead forecast in the forecasting period. Table \ref{t:25_quantileloss} reports the in-sample size for each series, which differs due to different non-trading days occurring in each market.

Two forecasting studies with different out-of-sample sizes are conducted. The first study aims to assess the performance of the models specifically for the 2008 GFC period, thus the initial date of the out-of-sample forecasting period is chosen as January 2008. Then for each index the out-of-sample size $m$ is chosen as 400, meaning that the end of the forecasting period is approximately falling around August 2009.

The second forecasting study incorporates a 8 year out-of-sample period, with the start date of the out-of-sample still chosen as Jan 2008 and out-of-sample size $m$ as 2000. Therefore, the end of the forecasting period is around the end of 2015.

Both daily one-step-ahead Value-at-Risk (VaR) and Expected Shortfall (ES) forecasts are considered for the returns on the 7 indices, using $\alpha= 2.5\%$, as recommended by \cite{BIS2019}.

For ES prediction, we have implemented the three weighted quantile approaches: WQ-Beta-M, WQ-EW-M and WQ-UNC-M, with $M\in\{3,5,10\}$. Similar to the simulation section, we have also considered the simple average with bias correction, SA-BC-M, and the simple average without bias correction, SA-No-BC-M. For the prediction of first stage quantile forecasts, two different regression specifications, CAViaR-SAV and CAViaR-AS, have been implemented. The estimation of 1$^{st}$-stage CAViaR models has been performed following the procedure described in Section \ref{model_section}.

Furthermore, the forecasting performances of the methods proposed in this paper have been compared with those yielded by other previously proposed approaches. Namely, the ES-CAViaR models of \cite{tayl2017} are also included in the study, again employing the CAViaR-SAV and CAViaR-AS models for the specification of the quantile regression component. These models are estimated following the suggestions from \cite{tayl2017}.

%For the candidate parameter vectors, the CAViaR parameters are optimized separately using quantile regression.

To make the models comparable, the CAViaR employed in our proposed weighted quantile and ES-CAViaR has used the same set up. Then, to assist the optimization in the estimation of ES-CAViaR models, the initial values of the parameters of the ES component are also selected by means of an additional random sampling procedure, following \cite{tayl2017}. When the ES-CAViaR-Add model of expression (\ref{es_caviar_add_model}) is used, $10^4$ candidate parameter vectors are incorporated. For the simpler ES-CAViaR-Mult Model (\ref{es_caviar_mult_model}), $10^3$ candidate parameter values are used.

In addition, the following models have also been included in the forecasting comparison: the conventional GARCH (\cite{bollerslev1986generalized}), EGARCH (\cite{nelson1991conditional}) and GJR-GARCH (\cite{glosten1993relation}), all with Student-t errors; the GARCH employing Hansen's skewed-t distribution (\cite{hansen1994autoregressive}); the CARE (using the size of the expectile level grid as 50) with Symmetric Absolute Value (CARE-SAV) and asymmetric specifications (CARE-AS). The GARCH-t, EGARCH-t and GJR-GARCH-t models are estimated using the Econometrics toolbox included in the Matlab 2019b release. The GARCH-Skew-t and CARE models are estimated by maximum likelihood using the Matlab code developed by the authors.

\subsection{Evaluation of forecasting performance: quantile loss}\label{quantile_loss_section}

One-step-ahead forecasts of VaR and ES are generated for each day in the forecast period for each data series.

The standard quantile loss function is employed to compare the models for VaR forecast accuracy. This choice is motivated considering that the standard quantile loss function is strictly consistent, i.e. the expected loss is a minimum at the true quantile series. Thus, the most accurate VaR forecasting model should produce the minimized aggregated quantile loss function, given as:
\begin{equation}\label{q_loss_score}
\sum_{t=n+1}^{n+m}(\alpha-I(r_t<Q_t))(r_t-Q_t)  \,\, ,
\end{equation}
where $n$ is the in-sample size, and $m$ is the out-of sample size with $m \in \{400, 2000\}$. $Q_{n+1},\ldots,Q_{n+m}$ is a series of quantile forecasts at level $\alpha=2.5\%$ for the observations $r_{n+1},\ldots,r_{n+m}$.

The quantile loss results are presented in Table \ref{t:25_quantileloss} for each model and for each series. The average loss is included in the ``Avg Loss'' column. The average rank based on ranks of quantile loss across 7 markets is calculated and shown in the ``Avg Rank'' column. Box indicates the favoured model and dashed box indicates the 2nd ranked model based on the average loss and rank.

It is worth reminding that, in the first stage of the weighted quantile estimations, VaR predictions are obtained via the estimation of either CAViaR-SAV or CAViaR-AS models, named as WQ-SAV or WQ-AS in Table \ref{t:25_quantileloss}. Depending on their ability to account for leverage effects in VaR dynamics, the tested models can be grouped into two categories: symmetric and asymmetric. For example, the GARCH-t, CARE-SAV, ES-CAViaR-Add-SAV, ES-CAViaR-Mult-SAV and WQ-SAV models have symmetric volatility or quantile (expectile) component, while the EGARCH-t, GJR-GARCH-t, CARE-AS, ES-CAViaR-Add-AS, ES-CAViaR-Mult-AS and WQ-AS have asymmetric ones.

%To have fair comparisons and more meaningful conclusions, the symmetric and asymmetric models need to be compared separately.

Based on the quantile loss results, we can see that the proposed weighted quantile and ES-CAViaR type models are characterized by very close performances, since the CAViaR components used in our proposed weighted quantile and ES-CAViaR models have the same set up, to make the next step ES comparison a fair one.

Also, in general and as expected, asymmetric models tend to perform slightly better than symmetric ones. For the SAV type models, the average quantile loss is around 58, while, for the AS type models, this average stays around 56, regarding the forecasting study on GFC period. For the study with longer forecasting horizon, the SAV and AS type models have average quantile loss around 172 and 167 respectively. The ES-CAViaR framework has the CAViaR parameters re-estimated when estimating ES, thus we observe minor quantile loss differences between the WQ and ES-CAViaR frameworks.

For both forecasting studies, EGARCH-t, GJR-GARCH-t and CARE-AS have slightly higher average quantile loss values, compared with ES-CAViaR-Add-AS, ES-CAViaR-Add-AS and WQ-AS type models. The symmetric GARCH-t and CARE-SAV have relatively less preferred performance compared with the ES-CAViaR-Add-SAV, ES-CAViaR-Mult-SAV and WQ-SAV models.

\begin{table}[hbt!]
\begin{center}
\caption{\label{t:25_quantileloss} \small 2.5\% quantile loss function values across the markets.}\tabcolsep=10pt
\tiny
\begin{tabular}{lccccccc|cc} \hline
Model&S\&P500&NASDAQ&HangSeng&FTSE&DAX&SMI&ASX200&Avg Loss&Avg Rank\\
\hline

GARCH-t&57.6&62.4&72.8&56.1&55.6&55.1&48.1&58.2&8.3\\
EGARCH-t&59.8&62.8&67.1&54.7&53.4&51.4&47.8&56.7&4.7\\
GJR-GARCH-t    &55.4&62.0&67.3&54.9&54.0&52.3&47.2&56.2&4.1\\
GARCH-Skew-t&56.5&61.8&72.6&55.3&54.4&54.9&47.0&57.5&5.7\\
CARE-SAV   &60.2&65.3&66.3&57.0&58.6&53.9&51.8&59.0&10.1\\
CARE-AS&61.1&66.8&61.7&55.0&55.9&54.9&48.9&57.7&8.7\\

ES-CAViaR-Add-SAV&59.6&63.8&68.5&55.7&56.8&53.3&48.1&58.0&8.6\\
ES-CAViaR-Mult-SAV&58.7&63.4&70.3&55.3&57.1&53.3&47.8&58.0&7.3\\

ES-CAViaR-Add-AS&60.6&63.0&63.2&52.9&53.7&51.4&47.0&\fbox{56.0}&\fbox{4.0}\\
ES-CAViaR-Mult-AS&59.1&63.7&65.0&52.8&54.4&51.6&46.0&\dbox{56.1}&4.3\\

WQ-SAV&59.5&63.0&69.5&55.6&56.9&53.3&48.1&58.0&8.0\\
WQ-AS&59.9&63.6&63.4&53.3&53.4&51.5&46.8&\fbox{56.0}&\dbox{4.1}\\

\hline
Out-of-sample $m$ (GFC) &400&400&400&400&400&400&400&&\\
In-sample $n$&1905&1892&1890&1943&1936&1930&1871&&\\

\hline
GARCH-t&168.3&187.2&205.3&160.4&188.8&164.4&145.2&174.2&11.3\\
EGARCH-t&166.9&184.0&194.9&155.2&181.6&159.3&140.3&168.9&4.7\\
GJR-GARCH-t    &162.9&183.1&196.4&156.4&182.9&159.4&141.7&169.0&4.7\\
GARCH-Skew-t&165.4&183.7&204.6&158.7&185.4&163.3&142.7&172.0&7.1\\
CARE-SAV   &168.8&185.9&200.0&159.4&188.1&165.7&146.7&173.5&10.6\\
CARE-AS&168.2&184.2&189.8&154.6&181.0&163.5&142.8&169.2&6.0\\

ES-CAViaR-Add-SAV&169.2&185.4&202.0&158.8&187.7&162.5&143.3&172.7&9.1\\
ES-CAViaR-Mult-SAV&168.1&184.9&203.1&158.5&187.7&162.9&143.5&172.7&8.6\\

ES-CAViaR-Add-AS&166.5&181.5&190.4&153.2&179.2&158.6&140.6&\fbox{167.1}&\dbox{2.6}\\
ES-CAViaR-Mult-AS&165.2&182.7&192.4&152.7&179.8&158.4&139.2&\dbox{167.2}&\fbox{2.1}\\

WQ-SAV&169.0&184.2&204.0&158.5&187.9&162.1&143.2&172.7&8.4\\
WQ-AS&167.0&182.5&190.5&153.2&179.0&158.9&139.8&167.3&2.7\\

\hline
Out-of-sample $m$ &2000&2000&2000&2000&2000&2000&2000&&\\
In-sample $n$&1905&1892&1890&1943&1936&1930&1871&&\\
\hline
\end{tabular}
\end{center}
\emph{Note}:\small  Box indicates the favoured model and dashed box indicates the 2nd ranked model based on the average loss and rank.
\end{table}

\subsection{Evaluation of forecasting performance: VaR and ES joint loss}

In this section, we assess the ability of the different models under comparison to forecast VaR and ES jointly. To this purpose, Table \ref{t:25_veloss} reports, for each model and data series, the value of the loss function in Eq. (\ref{es_caviar_log_score}) aggregated over the out-of-sample period: $\mathbf{S}= \sum_{t=n+1}^{n+m} S_{t}$, with $m \in \{400, 2000\}$. We use this to jointly compare the VaR and ES forecasts from all models, because the AL log-score in Eq. (\ref{es_caviar_log_score}) is a strictly consistent scoring function that is jointly minimized by the true VaR and ES series.

As mentioned in Section \ref{data_section}, for ES prediction, incorporating $M\in\{3,5,10\}$ we have implemented: WQ-Beta-M, WQ-EW-M,  WQ-Beta-M,  SA-BC-M (the simple average with bias correction), and SA-No-BC-M (the simple average without bias correction). By further incorporating the CAViaR-SAV and CAViaR-AS, we have 30 frameworks to be tested. Including the other 10 competing models, we have 40 models in total in Table \ref{t:25_veloss_gfc} and \ref{t:25_veloss}.

With respect to the forecasting study on GFC period as in Table \ref{t:25_veloss_gfc}, based on the average VaR and ES joint loss values the ES-CAViaR-Mult-AS produces the smallest average loss, followed by the GJR-GARCH-t and the proposed WQ-Beta-3-AS. The WQ-Beta-3-AS model on average ranks as the best, followed by the ES-CAViaR-Mult-AS. As discussed in Section \ref{quantile_loss_section}, we have employed the same CAViaR specification for weighted quantile and ES-CAViaR models. Therefore, the good performance of the proposed weighted quantile framework lends evidence on its validity in forecasting ES.

In addition, the WQ-AS models on average rank better and produce lower loss function values compared to EGARCH-t and GJR-GARCH-t models. Comparing the ES-CAViaR-SAV type models with the proposed weighted quantile frameworks, especially the WQ-Beta-SAV, we can still see that the WQ-SAV models, based on different numbers of grid points $M$, have lower average loss and rank better than ES-CAViaR-Add-SAV and similar performance compared with ES-CAViaR-Mult-SAV. On the other end, the CARE-SAV model on average produces the highest average joint loss. The SA-No-BC framework produces slightly smaller loss values and rank similar, compared to CARE-SAV.

Lastly, the weighted quantile framework has consistently improved performance than the SA-BC, which demonstrates the usefulness of the weighted average scheme. In addition, the performance of SA-BC is clearly better compared with SA-No-BC, demonstrating the effectiveness of the bias correction.

Regarding the forecasting study with out-of-sample size 2000 as in Table \ref{t:25_veloss}, based on the average joint loss,  the top 2 performing models are ES-CAViaR-Mult-AS and ES-CAViaR-Add-AS models. Based on the average rank, the top ranked models are ES-CAViaR-Mult-AS, SA-BC-3-AS and WQ-Beta-3-AS models. In addition, the WQ-AS type models, including WQ-Beta-AS, WQ-EW-AS and WQ-UNC-AS, clearly produce better performances compared to EGARCH-t, GJR-GARCH-t and CARE-AS. Comparing the ES-CAViaR-SAV type models with the WQ-SAV type models, we still observe the proposed WQ-SAV frameworks have improved performance compared to ES-CAViaR-Add-SAV and close performance compared to ES-CAViaR-Mult-SAV.

Finally, we would like to emphasize that the weighted quantile frameworks using $M=3$ can already generate very competitive performance, for both forecasting studies. Such results lend evidence on the fact the proposed weighted quantile framework can work effectively without having significantly increased computation cost, compared to other models. Compared with ES-CAViaR models, the WQ type models estimate and forecast ES nonparametrically, without assuming any ad-hoc relationship between the ES and VaR dynamics.

Lastly, we incorporate the Model Confidence Set (MCS) \citep{hansen2011model} to assess the statistical significance of differences in the values of the VaR and ES joint loss observed for various models under comparison, avoiding multiple testing biases. The MCS results are presented in Appendix \ref{mcs_section} and again support the proposed weighted quantile frameworks.

\begin{table}[hbt!]
\begin{center}
\caption{\label{t:25_veloss_gfc} \small With out-of-sample size 400 (GFC period), 2.5\% VaR and 2.5\% ES joint loss function values across the markets.}\tabcolsep=10pt
\tiny
\begin{tabular}{lccccccc|cc} \hline
Model&S\&P500&NASDAQ&HangSeng&FTSE&DAX&SMI&ASX200&Avg Loss&Avg Rank\\
\hline
GARCH-t&1102.0&1134.5&1185.2&1115.6&1114.6&1084.6&1044.2&1111.5&23.6\\
EGARCH-t&1114.2&1131.5&1138.7&1108.8&1068.2&1031.9&1030.3&1089.1&10.6\\
GJR-GARCH-t    &1078.6&1124.4&1143.2&1107.0&1082.6&1047.7&1027.9&\dbox{1087.4}&12.9\\
GARCH-Skew-t&1084.2&1125.1&1182.7&1094.8&1098.0&1069.9&1026.2&1097.3&17.0\\
CARE-SAV   &1135.4&1179.7&1143.8&1119.4&1133.7&1058.1&1101.0&1124.4&35.9\\
CARE-AS&1141.3&1182.3&1115.3&1120.8&1094.1&1063.2&1070.8&1112.5&30.9\\

ES-CAViaR-Add-SAV&1115.5&1146.7&1165.4&1105.6&1126.2&1060.9&1060.5&1111.5&27.3\\
ES-CAViaR-Mult-SAV&1110.8&1141.7&1168.5&1098.0&1123.3&1052.5&1049.2&1106.3&19.6\\

ES-CAViaR-Add-AS&1127.4&1135.4&1125.4&1085.0&1075.0&1034.2&1041.8&1089.2&13.6\\
ES-CAViaR-Mult-AS&1121.3&1142.6&1134.1&1078.5&1075.6&1028.3&1024.2&\fbox{1086.4}&\dbox{9.0}\\
\hline
WQ-Beta-3-SAV&1116.0&1141.0&1166.5&1104.7&1121.8&1055.1&1054.7&1108.5&20.9\\
WQ-EW-3-SAV&1118.2&1144.1&1167.3&1106.9&1122.5&1054.3&1055.4&1109.8&23.9\\
WQ-UNC-3-SAV&1118.2&1144.4&1167.4&1105.4&1122.1&1055.0&1056.7&1109.9&24.3\\
SA-BC-3-SAV&1117.3&1143.1&1170.6&1108.9&1123.2&1056.7&1055.2&1110.7&26.3\\
SA-NO-BC-3-SAV&1125.7&1154.1&1172.8&1115.1&1129.7&1062.7&1059.2&1117.0&34.9\\

WQ-Beta-5-SAV&1114.4&1144.3&1167.6&1106.8&1120.8&1055.7&1054.2&1109.1&22.1\\
WQ-EW-5-SAV&1117.0&1145.4&1167.7&1109.0&1121.4&1054.8&1052.0&1109.6&24.4\\
WQ-UNC-5-SAV&1116.3&1145.9&1168.1&1107.3&1121.6&1055.1&1051.5&1109.4&24.7\\
SA-BC-5-SAV&1118.0&1144.2&1171.3&1111.5&1121.8&1057.4&1053.1&1111.0&26.3\\
SA-NO-BC-5-SAV&1127.8&1155.6&1174.0&1118.7&1129.5&1064.2&1058.5&1118.3&36.3\\

WQ-Beta-10-SAV&1115.5&1145.8&1167.3&1105.6&1121.8&1055.1&1052.8&1109.1&23.3\\
WQ-EW-10-SAV&1115.5&1146.2&1167.3&1107.4&1122.2&1054.7&1049.0&1108.9&23.9\\
WQ-UNC-10-SAV&1114.6&1146.6&1167.1&1106.4&1121.9&1054.6&1051.3&1108.9&22.9\\
SA-BC-10-SAV&1117.4&1145.0&1171.1&1110.2&1122.2&1057.2&1049.6&1110.4&26.9\\
SA-NO-BC-10-SAV&1128.2&1156.8&1173.9&1117.8&1130.9&1064.3&1056.0&1118.3&36.7\\

WQ-Beta-3-AS&1120.1&1142.3&1121.2&1090.9&1074.4&1028.2&1034.8&\dbox{1087.4}&\fbox{7.9}\\
WQ-EW-3-AS&1130.7&1144.8&1120.6&1094.9&1075.8&1029.0&1033.7&1089.9&13.7\\
WQ-UNC-3-AS&1132.7&1144.5&1121.9&1093.9&1075.1&1029.3&1036.5&1090.6&13.9\\
SA-BC-3-AS&1120.5&1144.8&1121.7&1094.1&1073.3&1031.2&1034.5&1088.6&11.1\\
SA-NO-BC-3-AS&1127.3&1154.6&1122.9&1101.8&1078.6&1035.4&1040.1&1094.4&20.7\\

WQ-Beta-5-AS&1119.6&1146.6&1123.3&1088.7&1077.9&1029.0&1031.4&1088.1&12.9\\
WQ-EW-5-AS&1123.6&1146.3&1122.4&1090.8&1077.5&1029.0&1030.4&1088.6&13.0\\
WQ-UNC-5-AS&1124.4&1146.7&1123.3&1089.6&1078.4&1028.0&1032.9&1089.1&14.4\\
SA-BC-5-AS&1122.2&1145.4&1123.1&1090.5&1075.6&1032&1031.2&1088.6&13.0\\
SA-NO-BC-5-AS&1130.4&1156.0&1124.7&1098.6&1082.0&1037.6&1038.0&1095.3&22.4\\

WQ-Beta-10-AS&1120.1&1146.9&1122.6&1091.2&1076.1&1031.3&1036.6&1089.3&15.4\\
WQ-EW-10-AS&1122.0&1145.9&1122.1&1092.5&1075.9&1030.3&1036.2&1089.3&14.0\\
WQ-UNC-10-AS&1120.9&1145.9&1122.0&1092.1&1075.7&1030.1&1035.4&1088.9&13.0\\
SA-BC-10-AS&1122.1&1144.6&1122.6&1092.6&1074.5&1033.8&1037.6&1089.7&13.7\\
SA-NO-BC-10-AS&1130.9&1155.9&1124.2&1101.3&1081.2&1040.5&1045.6&1097.1&23.1\\

\hline
Out-of-sample $m$&400&400&400&400&400&400&400&&\\
%In-sample $n$&1905&1892&1890&1943&1936&1930&1871&&\\
\hline

\end{tabular}
\end{center}
\emph{Note}:\small  Box indicates the favoured model and dashed box indicates the 2nd ranked model, based on the average loss and rank. The horizontal line splits the proposed models and competing models from the literature.
\end{table}

\begin{table}[hbt!]
\begin{center}
\caption{\label{t:25_veloss} \small With out-of-sample size 2000, 2.5\% VaR and 2.5\% ES joint loss function values across the markets.}\tabcolsep=10pt
\tiny
\begin{tabular}{lccccccc|cc} \hline
Model&S\&P500&NASDAQ&HangSeng&FTSE&DAX&SMI&ASX200&Avg Loss&Avg Rank\\
\hline

GARCH-t&4324.8&4568.2&4700.7&4244.3&4658.4&4300.1&4071.6&4409.7&39.0\\
EGARCH-t&4290.6&4517.6&4586.9&4192.1&4585.2&4245.2&3990.4&4344.0&25.6\\
GJR-GARCH-t    &4239.2&4490.2&4601.7&4183.8&4601.4&4253.6&4009.1&4339.9&19.1\\
GARCH-Skew-t&4254.7&4491.0&4681.8&4194.0&4595.4&4245.3&4009.5&4353.1&22.6\\
CARE-SAV   &4304.8&4528.0&4664.2&4204.9&4606.2&4294.2&4080.8&4383.3&34.0\\
CARE-AS&4274.9&4470.5&4550.7&4158.5&4518.5&4250.7&4031.3&4322.2&19.6\\

ES-CAViaR-Add-SAV&4304.4&4505.2&4678.6&4199.8&4612.0&4246.3&4042.7&4369.9&33.4\\
ES-CAViaR-Mult-SAV&4289.8&4498.8&4677.3&4190.4&4609.0&4239.3&4042.4&4363.8&27.4\\

ES-CAViaR-Add-AS&4242.3&4439.2&4551.9&4131.8&4506.8&4192.1&3992.9&\dbox{4293.9}&7.6\\
ES-CAViaR-Mult-AS&4242.3&4451.2&4564.2&4117.8&4509.1&4188.3&3977.4&\fbox{4292.9}&\fbox{5.9}\\

\hline
WQ-Beta-3-SAV&4293.8&4490.8&4697.9&4193.8&4607.7&4240.1&4037.4&4365.9&27.6\\
WQ-EW-3-SAV&4294.6&4494.0&4698.5&4192.8&4609.3&4242.1&4036.6&4366.8&30.0\\
WQ-UNC-3-SAV&4295.1&4494.0&4699.1&4194.1&4608.1&4239.3&4039.0&4367.0&29.9\\
SA-BC-3-SAV&4295.3&4493.1&4703.8&4192.7&4609.5&4243.3&4033.0&4367.3&30.7\\
SA-NO-BC-3-SAV&4307.9&4511.5&4716.4&4202.5&4620.8&4260.6&4034.8&4379.2&36.1\\

WQ-Beta-5-SAV&4288.5&4493.1&4698.2&4191.8&4607.4&4237.5&4034.3&4364.4&24.7\\
WQ-EW-5-SAV&4291.1&4494.8&4698.2&4189.8&4609.4&4240.7&4031.2&4365.0&26.9\\
WQ-UNC-5-SAV&4291.1&4494.9&4698.8&4191.6&4608.7&4238.1&4032.7&4365.1&26.7\\
SA-BC-5-SAV&4293.7&4493.5&4703.7&4190.6&4609.3&4241.8&4029.2&4366.0&27.7\\
SA-NO-BC-5-SAV&4308.0&4512.7&4717.4&4201.8&4622.6&4261.3&4032.7&4379.5&36.0\\

WQ-Beta-10-SAV&4290.4&4495.4&4699.2&4188.1&4605.5&4239.8&4035.5&4364.9&26.4\\
WQ-EW-10-SAV&4290.6&4496.3&4699.0&4187.6&4607.0&4242.8&4030.9&4364.9&26.3\\
WQ-UNC-10-SAV&4287.9&4496.1&4699.4&4187.9&4605.9&4241.0&4034.4&4364.6&26.4\\
SA-BC-10-SAV&4294.4&4495.2&4704.4&4188.5&4606.3&4244.5&4028.3&4366.0&28.0\\
SA-NO-BC-10-SAV&4310.3&4516.1&4718.3&4200.6&4621.0&4265.7&4033.3&4380.8&36.7\\

WQ-Beta-3-AS&4252.6&4449.0&4549.4&4130.4&4507.7&4195.2&3982.4&4295.2&\dbox{6.1}\\
WQ-EW-3-AS&4263.5&4451.5&4546.8&4131.0&4508.9&4196.1&3980.1&4296.8&7.9\\
WQ-UNC-3-AS&4264.4&4451.0&4549.0&4132.9&4509.0&4198.0&3982.9&4298.2&9.9\\
SA-BC-3-AS&4253.7&4450.9&4547.8&4130.0&4505.7&4196.8&3981.7&4295.2&\fbox{5.9}\\
SA-NO-BC-3-AS&4265.5&4467.9&4550.9&4145.5&4519.2&4214.4&3988.1&4307.4&15.1\\

WQ-Beta-5-AS&4255.0&4454.9&4550.9&4135.6&4513.3&4195.2&3979.4&4297.8&10.0\\
WQ-EW-5-AS&4257.5&4455.3&4548.4&4133.9&4512.0&4194.2&3976.8&4296.9&8.1\\
WQ-UNC-5-AS&4255.3&4454.9&4550.8&4135.7&4514.1&4194.2&3980.1&4297.9&10.0\\
SA-BC-5-AS&4261.2&4453.9&4549.2&4132.3&4509.3&4196.1&3976.6&4297.0&8.4\\
SA-NO-BC-5-AS&4275.3&4472.7&4553.0&4147.5&4525.4&4218.2&3984.9&4311.0&16.9\\

WQ-Beta-10-AS&4254.7&4454.3&4551.5&4129.7&4506.6&4195.4&3983.3&4296.5&8.6\\
WQ-EW-10-AS&4257.9&4454.5&4549.8&4129.4&4506.4&4195.7&3981.3&4296.4&7.6\\
WQ-UNC-10-AS&4253.3&4452.6&4550.2&4129.3&4506.8&4196.1&3981.6&4295.7&6.7\\
SA-BC-10-AS&4259.3&4452.7&4550.3&4128.3&4504.2&4198.2&3980.6&4296.2&7.7\\
SA-NO-BC-10-AS&4274.7&4473.4&4554.8&4143.3&4520.6&4222.7&3990.0&4311.4&16.9\\

\hline
Out-of-sample $m$&2000&2000&2000&2000&2000&2000&2000&&\\
%In-sample $n$&1905&1892&1890&1943&1936&1930&1871&&\\
\hline
\end{tabular}
\end{center}
\emph{Note}:\small  Box indicates the favoured model and dashed box indicates the 2nd ranked model, based on the average loss and rank. The horizontal line splits the proposed models and competing models from the literature.
\end{table}

\newpage
{\centering
\section{\normalsize CONCLUSION}\label{conclusion_section}
\par
}

In this paper, we propose an innovative semi-parametric weighted quantile framework for ES estimation and forecasting. The proposed approach relies on a two-step estimation procedure. The quantiles weighting scheme is parsimoniously parameterized by incorporating a Beta weight function whose coefficients are optimized by minimizing a strictly consistent joint VaR and ES loss function. Some alternative specifications of the quantile weighting function are also considered as robustness checks.
Through simulation study, we have demonstrated the effectiveness of the proposed framework. In an empirical study, focusing on the highly volatile GFC period, improvements in the out-of-sample forecasting of ES are observed, compared to traditional GARCH and CARE models, as well as the ES-CAViaR models. Empirical evidence on a longer forecasting period confirms the superiority of the WQ framework over GARCH-type and CARE models and its competitiveness with state-of-the-art approaches such as the ES-CAViaR models.

The proposed framework can be extended in a number of ways. First, at the moment the first stage of the framework only uses quantile estimates from CAViaR. However, the proposed framework is quite flexible, so it can actually incorporate quantile estimates obtained from any models. Second, during the second stage of the estimation (when estimating the parameters of Beta weight function), we can also re-estimate the CAViaR parameters to potentially further improve the VaR and ES estimation accuracy, similar to the ES-CAViaR estimation. Third, the framework can be also extended by allowing different quantiles to be estimated with different models (not limited to CAViaR). Then in order to generate the quantile forecast for each quantile level, we can select the model by cross validation or use model combination, see \cite{taylor2020forecast} as example.

%\clearpage
%\appendixtitleon
%\appendixtitletocon
%\begin{appendices}

%{\centering
%\section{\normalsize Derivation details}\label{derivation}
%\par
%}

%\bigskip
%{\centering
%\section{\normalsize SIMULATION RESULTS}\label{simulation_results_section}
%\par
%}

%\end{appendices}

%\clearpage

\bibliographystyle{chicago}
\bibliography{bibliography}

\clearpage
\appendixtitleon
\appendixtitletocon
\begin{appendices}

{\centering
\section{Beta weight function}\label{beta_function_section}
\par
}
\setcounter{figure}{0} \renewcommand{\thefigure}{A.\arabic{figure}}
\setcounter{table}{0} \renewcommand{\thetable}{A.\arabic{table}}

\begin{figure}[htp]
    \centering
    \includegraphics[width=0.35\textwidth]{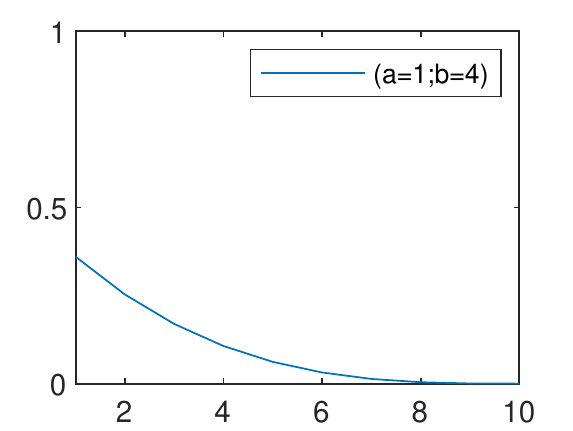}
    \includegraphics[width=0.35\textwidth]{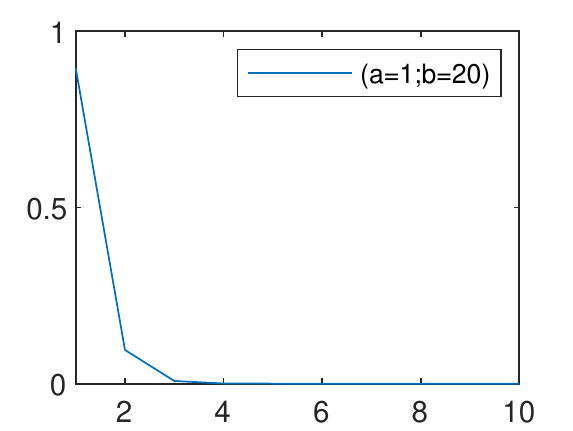}
    \includegraphics[width=0.35\textwidth]{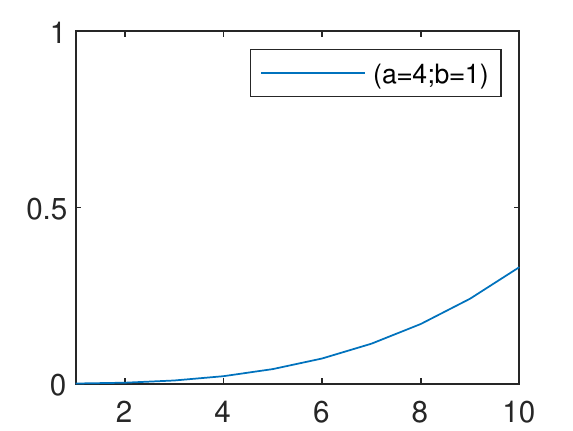}
    \includegraphics[width=0.35\textwidth]{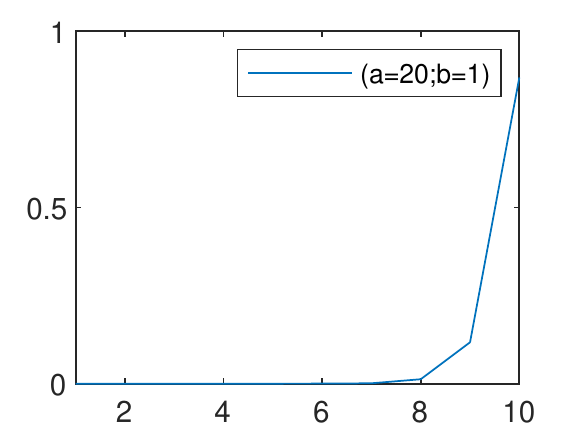}
        \includegraphics[width=0.35\textwidth]{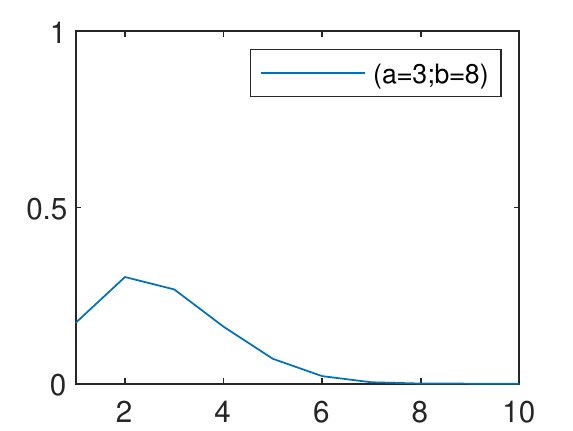}
    \includegraphics[width=0.35\textwidth]{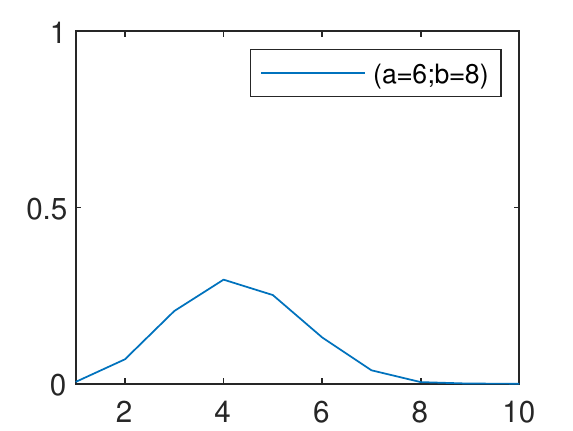}
        \includegraphics[width=0.35\textwidth]{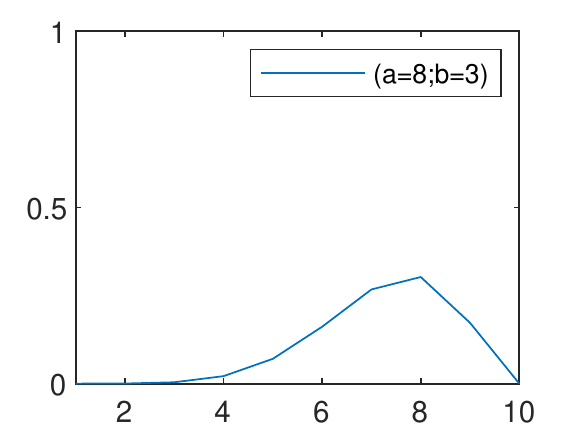}
    \includegraphics[width=0.35\textwidth]{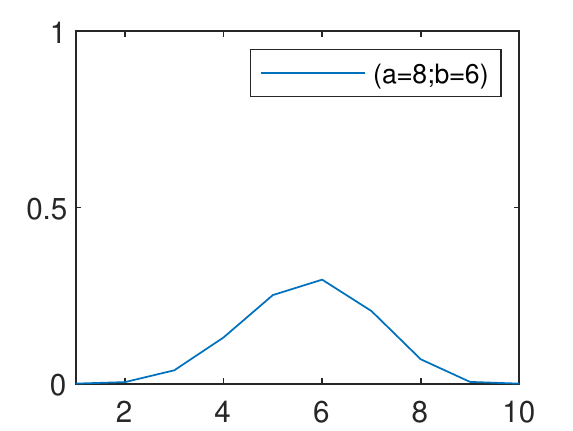}
    \caption{The figure displays various weighting patterns generated by the Beta weight function for different values of the parameters $a$ and $b$.
    % (from left to right and from top to bottom): ($a=1$,$b=4$), ($a=1$, $b=20$), ($a=4$, $b=1$), ($a=20$, $b=1$), ($a=3$, $b=8$), ($a=6$, $b=8$), ($a=8$, $b=3$), ($a=8$, $b=6$).
    }
    \label{f:betapatt}
\end{figure}

For different values of the coefficients $a$ and $b$ in Eq. (\ref{e:betalag}), Fig. \ref{f:betapatt} displays various patterns that can be generated under the Beta weight structure. To facilitate comparison among different patterns, the weights in the plots have been normalized so that they sum up to unity.
\captionsetup[figure]{labelformat={default},labelsep=period,name={Fig.}}

{\centering
\section{Model confidence set}\label{mcs_section}
\par
}
\setcounter{figure}{0} \renewcommand{\thefigure}{B.\arabic{figure}}
\setcounter{table}{0} \renewcommand{\thetable}{B.\arabic{table}}

The MCS \citep{hansen2011model} is used to assess the statistical significance of differences in the values of the VaR and ES joint loss produced from all competing models.

A MCS is a set of models that is constructed such that it will contain the best model with a given level of confidence (75\% is used in our paper). All computations are performed using the Matlab code for MCS testing included in Kevin Sheppard's MFE toolbox\footnote{The code can be downloaded at the url ``https://www.kevinsheppard.com/code/matlab/mfe-toolbox/''}. The R and SQ methods which use absolute and squared values sum respectively during the calculation of test statistic are employed in our paper, details as in page 465 of \citep{hansen2011model}.

Table \ref{mcs_r} presents the 75\% MCS using both the R and SQ methods, for two forecasting studies with out-of-sample size $m$ as 400 and 2000 respectively. Columns ``R-Total-GFC'' and ``SQ-Total-GFC'', ``R-Total'' and ``SQ-Total'' count the total number of times that a model is included in the 75\% MCS across the 7 return series.

Overall, we observe our weighted quantile models are more or equally likely to be included in MCS, in comparison with other models. For both R and SQ methods across two forecasting studies, in addition to ES-CAViaR-Mult-AS, most of the weighted quantile models with AS specification are included in MCS for all 7 series. Regarding the weighted quantile models with SAV specification, we still observe that the weighted quantile models are in general more or equally to be included in the MCS, comparing to GARCH-t, CARE-SAV or ES-CAViaR-SAV.

\begin{table}[hbt!]
\begin{center}
\caption{\label{mcs_r} \small 75\% model confidence set results summary with R and SQ methods.}\tabcolsep=10pt
\tiny
\begin{tabular}{lcc|ccccc} \hline
Model&R-Total-GFC&SQ-Total-GFC&R-Total&SQ-Total\\
\hline

GARCH-t&5&4&3&3\\
EGARCH-t&\dbox{6}&\fbox{7}&\dbox{6}&\dbox{6}\\
GJR-GARCH-t    &\dbox{6}&\fbox{7}&\dbox{6}&\dbox{6}\\
GARCH-Skew-t&\dbox{6}&5&5&5\\
CARE-SAV   &5&\dbox{6}&3&3\\
CARE-AS&5&\dbox{6}&\fbox{7}&\fbox{7}\\

ES-CAViaR-Add-SAV&4&4&4&4\\
ES-CAViaR-Mult-SAV&\dbox{6}&5&4&4\\

ES-CAViaR-Add-AS&\dbox{6}&\fbox{7}&\fbox{7}&\fbox{7}\\
ES-CAViaR-Mult-AS&\fbox{7}&\fbox{7}&\fbox{7}&\fbox{7}\\
\hline
WQ-Beta-3-SAV&5&5&5&4\\
WQ-EW-3-SAV&4&5&5&4\\
WQ-UNC-3-SAV&5&5&4&4\\
SA-BC-3-SAV&4&5&5&4\\
SA-NO-BC-3-SAV&4&4&4&4\\

WQ-Beta-5-SAV&4&5&5&4\\
WQ-EW-5-SAV&4&5&5&4\\
WQ-UNC-5-SAV&4&5&5&4\\
SA-BC-5-SAV&4&5&5&4\\
SA-NO-BC-5-SAV&3&4&4&4\\

WQ-Beta-10-SAV&5&5&5&4\\
WQ-EW-10-SAV&4&5&5&4\\
WQ-UNC-10-SAV&4&5&5&4\\
SA-BC-10-SAV&4&5&5&4\\
SA-NO-BC-10-SAV&3&3&4&4\\

WQ-Beta-3-AS&\fbox{7}&\fbox{7}&\fbox{7}&\fbox{7}\\
WQ-EW-3-AS&\fbox{7}&\fbox{7}&\fbox{7}&\fbox{7}\\
WQ-UNC-3-AS&\fbox{7}&\fbox{7}&\fbox{7}&\fbox{7}\\
SA-BC-3-AS&\fbox{7}&\fbox{7}&\fbox{7}&\fbox{7}\\
SA-NO-BC-3-AS&5&\fbox{7}&\fbox{7}&\dbox{6}\\

WQ-Beta-5-AS&\fbox{7}&\fbox{7}&\fbox{7}&\fbox{7}\\
WQ-EW-5-AS&\fbox{7}&\fbox{7}&\fbox{7}&\fbox{7}\\
WQ-UNC-5-AS&\fbox{7}&\fbox{7}&\dbox{6}&\dbox{6}\\
SA-BC-5-AS&\fbox{7}&\fbox{7}&\fbox{7}&\fbox{7}\\
SA-NO-BC-5-AS&4&\dbox{6}&\dbox{6}&5\\

WQ-Beta-10-AS&\fbox{7}&\fbox{7}&\fbox{7}&\fbox{7}\\
WQ-EW-10-AS&\fbox{7}&\fbox{7}&\fbox{7}&\fbox{7}\\
WQ-UNC-10-AS&\fbox{7}&\fbox{7}&\fbox{7}&\fbox{7}\\
SA-BC-10-AS&\fbox{7}&\fbox{7}&\fbox{7}&\fbox{7}\\
SA-NO-BC-10-AS&4&\dbox{6}&\fbox{7}&5\\

\hline
Out-of-sample $m$ &400&400&2000&2000\\
\hline
\end{tabular}
\end{center}
\emph{Note}:\small Boxes indicate the favoured model and dashed box indicates the 2nd ranked model, based on the number of times that a model is included in the MCS across the 7 markets (higher is better). The horizontal line splits the proposed models and competing models from the literature.
\end{table}

\end{appendices}

\end{document}